\documentclass[11pt,british]{article}
\PassOptionsToPackage{linecolor=blue,backgroundcolor=blue!25,bordercolor=blue,textsize=scriptsize}{todonotes}
\PassOptionsToPackage{dvipsnames}{xcolor}
\usepackage[utf8]{inputenc}
\usepackage[letterpaper]{geometry}
\usepackage{color}
\usepackage{babel}
\usepackage{textcomp}
\usepackage{mathtools}
\usepackage{todonotes}
\usepackage{booktabs}  
\usepackage{amsmath}
\usepackage{amsthm}
\usepackage{amssymb}
\usepackage{stmaryrd}
\SetSymbolFont{stmry}{bold}{U}{stmry}{m}{n}
\usepackage{setspace}
\usepackage{microtype}
\usepackage{soul}
\usepackage{graphicx}
\usepackage[unicode=true,
 bookmarks=true,bookmarksnumbered=false,bookmarksopen=false,
 breaklinks=false,pdfborder={0 0 0},pdfborderstyle={},backref=false,colorlinks=true]
 {hyperref}
\hypersetup{
colorlinks=true,urlcolor=darkblue,anchorcolor=darkblue,citecolor=darkblue,filecolor=darkblue,linkcolor=darkblue,menucolor=darkblue,linktocpage=true,pdftitle = {Deep learning lattice gauge theories},}

\makeatletter

\usepackage{tikz}
\usetikzlibrary{spy}
\usetikzlibrary{backgrounds}
\usepackage[framemethod=tikz]{mdframed}

\usepackage{slashed}	 	%
\usepackage{amsfonts} 		%
\usepackage{cite}	%
\SetTracking{encoding={*}, shape=sc}{0} 	%
\allowdisplaybreaks		%
\numberwithin{equation}{section}	%
\usepackage[title,header]{appendix}

\makeatletter
\g@addto@macro\bfseries{\boldmath}
\makeatother

\usepackage{xcolor}
\definecolor{darkblue}{rgb}{0.2,0.0,0.6}

\makeatletter
\g@addto@macro\@floatboxreset\centering
\makeatother

\let\originalleft\left
\let\originalright\right
\renewcommand{\left}{\mathopen{}\mathclose\bgroup\originalleft}
\renewcommand{\right}{\aftergroup\egroup\originalright}

\usepackage{pgfplots}
\usepgfplotslibrary{colorbrewer}

\pgfplotsset{
    fullplot/.style={
    tick style={thin},
    grid style={dashed,gray}, 
    axis lines=center,
    width=9cm,
    height=6.3cm,
    scale only axis,
    major tick style = black,
    cycle list/Dark2-8,
    grid=both,
    legend style={font=\small},
    legend cell align={left},
    mark options={draw opacity=1},  %
    x label style={anchor=west},
    y label style={anchor=south},
    x tick label style={align=center,text width=2cm, font=\small},
    y tick label style={font=\small},
    nodes near coords align={vertical},
	grid=both},
    every axis plot/.append style={smooth, thick, tension=0.2, draw opacity = 0.5},
}

\pgfplotsset{
    halfplot/.style={
    fullplot,    
    width=6.5cm,
    height=4.55cm,
    legend style={font=\footnotesize,},
    x tick label style={align=center,text width=2cm, font=\footnotesize},
    y tick label style={font=\footnotesize}},
}

\tikzstyle{trimmed plot}=[tight background, trim axis left, trim axis right]

\usetikzlibrary{cd}
\usepgfplotslibrary{external}
\pgfplotsset{compat=1.18}

\makeatother

\begin{document}
\global\long\def\ii{\text{i}}%
\global\long\def\ee{\text{e}}%
\global\long\def\Orth#1{\text{O}(#1)}%
\global\long\def\Uni#1{\text{U}(#1)}%
\global\long\def\SU#1{\text{SU}(#1)}%
\global\long\def\op#1{\operatorname{#1}}%
\global\long\def\im{\operatorname{im}}%
\global\long\def\re{\operatorname{re}}%
\global\long\def\eqspace{\mathrel{\phantom{{=}}{}}}%
\global\long\def\bZ{\mathbb{Z}}%
\global\long\def\bC{\mathbb{C}}%
\global\long\def\bP{\mathbb{P}}%
\global\long\def\bR{\mathbb{R}}%
\global\long\def\gc{g_\text{c}}%
\global\long\def\id{\boldsymbol{1}}%

\begin{titlepage}
\begin{flushright} %
\end{flushright}
\vfill
\begin{center}
{\setstretch{1.5}\LARGE\bf Deep learning lattice gauge theories\par} 
\vskip 1cm 
Anuj Apte,\textsuperscript{a} Anthony Ashmore,\textsuperscript{b,c} Clay Córdova,\textsuperscript{a} and Tzu-Chen Huang\textsuperscript{a} 
\vskip 1cm
\textit{\small\textsuperscript{a}Enrico Fermi Institute \& Kadanoff Center for Theoretical Physics,\\ University of Chicago, Chicago, IL 60637, USA}
\\[.25cm]
\textit{\small\textsuperscript{b}Laboratoire de Physique Théorique et Hautes \'Energies,\\ Sorbonne Universit\'e, UPMC Paris 06, UMR 7589, 75005 Paris, France}
\\[.25cm]
\textit{\small\textsuperscript{c}Department of Physics, Skidmore College,\\
 Saratoga Springs, NY 12866, USA}

\end{center}
\vfill
\begin{center} \textbf{Abstract} \end{center}
\begin{quote}
Monte Carlo methods have led to profound insights into the strong-coupling behaviour of lattice gauge theories and produced remarkable results such as first-principles computations of hadron masses. Despite tremendous progress over the last four decades, fundamental challenges such as the sign problem and the inability to simulate real-time dynamics remain. Neural network quantum states have emerged as an alternative method that seeks to overcome these challenges. In this work, we use gauge-invariant neural network quantum states to accurately compute the ground state of $\mathbb{Z}_N$ lattice gauge theories in $2+1$ dimensions. Using transfer learning, we study the distinct topological phases and the confinement phase transition of these theories. For $\mathbb{Z}_2$, we identify a continuous transition and compute critical exponents, finding excellent agreement with existing numerics for the expected Ising universality class. In the $\mathbb{Z}_3$ case, we observe a weakly first-order transition and identify the critical coupling. Our findings suggest that neural network quantum states are a promising method for precise studies of lattice gauge theory.
\end{quote}
\vfill
{\begin{NoHyper}\let\thefootnote\relax\footnotetext{\tt \!\!\!\!\!\!\!\!\!\!\! apteanuj@uchicago.edu, aashmore@skidmore.edu, clayc@uchicago.edu, tzuchen@uchicago.edu}\end{NoHyper}}
\end{titlepage}

\tableofcontents

\section{Introduction}

Since their introduction by Wilson~\cite{Wilson:1974sk, Kogut:1974ag}, lattice gauge theories have been pivotal in understanding fundamental physics, particularly quantum chromodynamics (QCD)~\cite{Wilczek2000, Greiner2006-da}, which governs the strong nuclear force. QCD, a non-Abelian gauge theory, describes interactions between quarks and gluons, the elementary constituents of hadrons. Despite its apparent simplicity, the non-perturbative regime of this theory exhibits remarkably complex phenomena, including confinement~\cite{Greensite2011} and chiral symmetry breaking~\cite{Nambu:1961tp, Nambu:1961fr}, often resisting analytical tools. As such, Monte Carlo-based numerical methods have become invaluable~\cite{Rebbi1983}, providing insights into QCD phases~\cite{Fodor:2001pe} and enabling first-principles calculations of hadron masses~\cite{BMW:2008jgk}.

Discrete lattice gauge theories -- lattice gauge theories with \emph{finite} gauge groups -- serve as important models for studying the non-perturbative behaviour and emergent phenomena of gauge theories. While simpler than their continuous counterparts, they still capture key features such as gauge invariance, confinement and topological excitations. This makes them valuable testing grounds for developing and benchmarking new theoretical and computational techniques.

The present work focuses on finding the ground-state wavefunction of discrete lattice gauge theories in $2+1$ dimensions. Since the wavefunction contains rich information about quantum correlations and topological properties of the system, there are many reasons to directly compute the wavefunction itself instead of simply calculating expectation values of observables via traditional (Euclidean) Monte Carlo. For example, entanglement measures such as Rényi entropies can be computed with this approach, since the wavefunction encodes the full correlation structure of the lattice system~\cite{Vishwanath2011,Tubman2013}. 

Obtaining accurate ground-state wavefunctions for lattice gauge theories is challenging due to the complexity of the non-perturbative regime and the exponential growth of the Hilbert space with system size. In particular, traditional analytical and numerical methods often struggle to capture the intricate correlations and entanglement present in the ground state of these systems. Recent years have shown that deep neural networks are adept at learning and sampling from classical probability distributions across a variety of domains, including text, images and speech~\cite{LeCun2015}. The problem of finding a ground-state wavefunction can be viewed as a generalization of this problem, namely, computing the complex-valued probability amplitude associated to the basis elements of the underlying Hilbert space, so it is not surprising that neural networks are well suited to this task.

The use of neural networks to represent wavefunctions was pioneered by Carleo and Troyer~\cite{1606.02318}, and has since become known as a ``neural network quantum state'' (NQS or NNQS). The idea behind a neural network quantum state is to model the wavefunction as a deep neural network with hidden layers, leveraging the expressive power of these networks to construct a highly flexible variational ansatz for the ground state. One then ``trains'' the network to minimise the energy of the variational state, with expectation values calculated using Markov chain Monte Carlo. Key to this process is the ability to efficiently compute gradients of expectation values with respect to the parameters of the network, so that the energy can be minimised via (stochastic) gradient descent~\cite{Chong2013-ot}. Thanks to differentiable programming and backpropagation~\cite{JMLR:v18:17-468,Rumelhart1986}, one can compute these gradients to machine precision without resorting to numerical interpolation. At the end of the training process, one hopes to have a NNQS that can accurately capture the physics of the ground state and enable calculation of other observables. 

There are two fundamental problems in simulating quantum many-body systems: the problems of representation and computation. Focusing on a discrete lattice system for concreteness, a wavefunction associates a complex number to every lattice configuration. Since the number of configurations grows exponentially with the system size, naively one might need an ansatz with a similarly large number of parameters. 
This is the problem of representation. There is no reason to expect that a randomly chosen state of the full Hilbert space can be represented efficiently, but fortunately physical states -- and ground states in particular -- are not generic, instead exhibiting features such as area-law entanglement~\cite{Hastings2007}. NNQSs provide an efficient and flexible parametrisation of these states with a sub-exponential number of variational parameters.\footnote{NNQSs should also be compared with tensor networks for $1+1$-dimensional systems. Tensor networks are known to be complete (any pure state can be described by increasing the bond dimension sufficiently) and efficient for states with area-law entanglement (the cost of computing expectation values grows polynomially with the number of parameters). Similarly, thanks to universal representation theorems~\cite{Cybenko:1989iql}, NNQSs can describe arbitrary quantum states, including ground states with volume-law entanglement~\cite{Denis:2023dww}. Although there is no bound on the size of the network required to encode these states, it is known that NNQSs can exactly reproduce tensor network or projected entangled pair states, and there are examples where NNQSs provide efficient representations of physically relevant states where these other approaches fail to be efficient~\cite{1701.04831,Sharir:2021dpe}.} Given such an efficient encoding of a quantum state, one still needs to be able to compute probability amplitudes in a reasonable time. Furthermore, physical observables are still operators on an exponentially large Hilbert space, so computing expectation values may be challenging. These comprise the problem of computation. For NNQSs, it is simple and fast to compute the amplitude for an input lattice configuration, with observables then calculated by restricting to a finite sum over Hilbert space via Monte Carlo importance sampling~\cite{Weinzierl:2000wd}.

Much like traditional lattice Monte Carlo, NNQSs can be used to study finite-temperature systems~\cite{2103.04791}. There are also a number of questions that are difficult to tackle with Euclidean methods but which are within reach of NNQSs, including real-time and dissipative dynamics, fermionic systems with finite chemical potential, quantum-state reconstruction, and open quantum systems~\cite{Lange:2024nsr}. It is possible to impose both global and gauge symmetries exactly in NNQS by using invariant neural networks~\cite{Medvidovic:2024ihh}. 

The particular neural network architecture that we employ is based on ``lattice gauge-equivariant convolutional neural networks'' (L-CNNs), introduced by Favoni et al.~\cite{Favoni:2020reg}. L-CNNs possess the usual locality and weight sharing properties of traditional convolutional networks. Crucially, they are also manifestly gauge equivariant and so one does not have to impose gauge symmetry via inexact methods, such as energy penalties. Their uses to date have focused on predicting gauge-invariant observables via supervised learning~\cite{Favoni:2021epq,Favoni:2021flp,Favoni:2022mcg,Aronsson:2023rli,Wenger:2023gem,Holland:2024muu}. We instead use an L-CNN as a NNQS, i.e.~a variational ansatz for the ground-state wavefunction, similar to the approach taken by Luo et al.~\cite{Luo:2020stn,Luo:2022jzl}. We have implemented this architecture using NetKet~\cite{Vicentini:2021pcv}, a flexible machine-learning framework for many-body quantum physics built on top of JAX~\cite{JAX}, enabling automatic differentiation and GPU acceleration.

\subsubsection{Previous work}

Machine learning methods have already made an impact in many areas of physics, including astronomy and cosmology, condensed-matter physics, fluid dynamics, quantum chemistry, and nuclear and particle physics, and have shown potential for studying lattice gauge theories~\cite{Boyda:2022nmh}.

Since our approach is based on an existing network architecture, it is useful to outline how our work adds to the state of the art. Favoni et al.~\cite{Favoni:2020reg} introduced L-CNNs to incorporate gauge symmetries directly into the network structure while maintaining the ability to approximate any gauge-covariant function on the lattice. The authors showed that L-CNNs outperform conventional CNNs in regression tasks involving Wilson loops of different sizes and shapes in pure $\SU2$ gauge theory, with the performance gap increasing as the loop size grows. This supervised learning problem demonstrated that L-CNNs can learn the physical information encoded in Wilson loops and other gauge-invariant observables. There have been a number of follow-up works which explore the generalization ability of L-CNNs and their geometric interpretation~\cite{Bulusu:2021rqz,Aronsson:2023rli}, but no attempt has been made to use this architecture as a variational ansatz for the ground state of a lattice theory.

Closer to our approach is the work by Luo et al.~\cite{Luo:2020stn}, where gauge-equivariant neural network quantum states were introduced as a technique to efficiently model the wavefunction of lattice gauge theories while maintaining exact local gauge invariance. For the specific case of $\bZ_2$ gauge theory on a square lattice, the authors combined this architecture with variational Monte Carlo to study the ground state and the confinement-deconfinement transition. This was extended to include matter in \cite{Luo:2021ccm} and to $\Uni1$ gauge theory in \cite{Luo:2022jzl}. The aims of the present paper include greatly improving upon the precision of these results and studying $\bZ_3$ gauge theory. For example, these studies did not determine the order of the phase transition, calculate the critical exponents, investigate static charges, or provide a precise value for the critical coupling. In filling these gaps, our goal is to demonstrate that deep learning can be used for precise calculation, and in doing so enhance our understanding of the phases and phase transitions in discrete gauge theories.

\subsubsection{Main results}

For $\bZ_2$ lattice gauge theory, we compute the ground-state energy as a function of coupling for various lattice sizes (Figure \ref{fig:Z2_LxL_energy}) and confirm the existence of confined and deconfined phases via calculations of order and disorder parameters and the string tension. We show that the confinement phase transition between the two phases is continuous, with a critical coupling and critical exponents given by
\begin{equation}
    \gc = 0.7546(8),\qquad\beta = 0.326(4), \qquad\nu = 0.630(3).
\end{equation}
These values are calculated via ``curve collapse'' from finite-scaling theory. The high quality of the resulting collapse can be seen in Figure \ref{fig:tHooft_collapse}. These values imply that the conformal dimensions of the $\sigma$ and $\epsilon$ primary operators in the $2+1$-d Ising CFT are:
\begin{equation}
    \Delta_\sigma = 0.518(4), \qquad\Delta_\epsilon = 1.412(8),
\end{equation}
in excellent agreement with previous results from perturbative calculations, Monte Carlo simulations and the conformal bootstrap. We further refine our estimate of the critical coupling using BST extrapolation~\cite{10.1007/BF02165234, 10.1088/0305-4470/20/8/032, 10.1088/0305-4470/21/11/019} and find $\gc=0.756431$. We also confirm the linear potential between static charges and compute it as a function of coupling, finding it vanishes for small couplings and smoothly increases around the critical coupling, approaching the strong-coupling prediction for large couplings. 

For $\bZ_3$, we compute the ground-state energies for varying lattice size (Figure \ref{fig:Z3_LxL_energy}) and, for small lattices, confirm that they agree with results from exact diagonalization. As with $\bZ_2$, we show the existence of confined and deconfined phases via Wilson loop and `t Hooft string operators, and the string tension. We investigate the nature of the phase transition by examining the electric and magnetic energies, and derivatives of the energy as a function of coupling. We plot these in Figure \ref{fig:Z2_Z3_E_deriv}, finding that $\bZ_3$ displays a sharp change at the transition compared with a smooth change for $\bZ_2$, indicating the presence of a first-order transition. This transition occurs within the range $g\in[0.775,0.7875]$, giving a critical coupling of
\begin{equation}
    \gc = 0.781(6).
\end{equation}
Due to finite system size and the weakly first-order nature of the transition, we do not observe a discontinuity in either the energy derivatives or the order and disorder parameters.

We also quantify how well our networks have learned the charge conjugation symmetry of the $\bZ_3$ theory by measuring the imaginary component of Wilson loop expectations, which should vanish for the exact ground state. Depending on the value of the coupling, we find that this holds to better than one part in ${10}^4$ to ${10}^6$. Finally, we compute the static potential between charges, recovering the expected linear potential and finding agreement with perturbative results in the small and large coupling regimes.

\begin{figure*}
\hfill{}
\includegraphics[trim={32.10236pt 0 15.28946pt 0}]{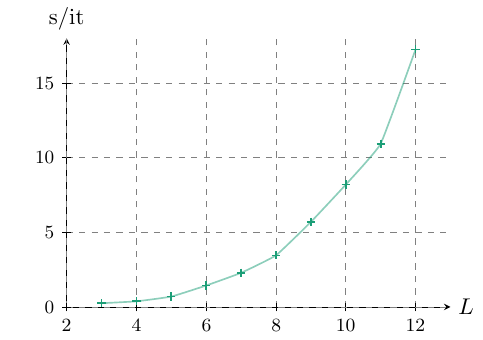}%
\hfill{}
\includegraphics[trim={32.10236pt 0 15.28946pt 0}]{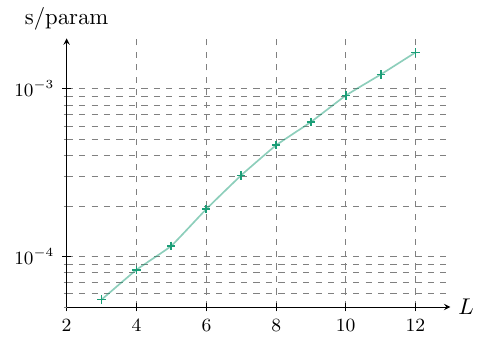}%
\hfill{}
\caption{The left figure shows seconds per training iteration as a function of lattice size $L$. The depths of the networks are chosen so that, with kernel size $K=2$ per layer, the largest Wilson loop is large enough to cover the entire lattice. All other hyperparameters are identical. The right figure shows the number of seconds per network parameter for a single iteration. The computations were performed on a single NVIDIA A100 GPU.}\label{fig:timing}
\end{figure*}

The NNQSs that we use to obtain these results have up to 10k parameters, depending on the depth and width of the networks. To illustrate the computational resources and time needed for training, in Figure \ref{fig:timing} we plot the number of seconds per iteration and the number of seconds per network parameter for a single iteration for various lattice sizes. The networks are the same size as those used to obtain the results in the main text, with $(N_\text{out},K)=(4,2)$ (see Section \ref{sec:LCNN}) and depths given by $\bigl\lceil \log_2 L^2 \bigr\rceil$ for an $L\times L$ lattice, so that the largest Wilson loops constructed by the network can cover the full lattice. We train our networks for 500 to 1000 iterations, with smaller networks converging well before this and larger networks benefitting from longer training. The networks used for studying $L=12$ systems train in approximately five hours on a single NVIDIA A100 GPU~\cite{Krashinsky2021}. In fact, the networks often converge well before this if one utilises transfer learning when scanning over couplings (see Section \ref{sec:netket}).

\subsubsection{Outline}

We begin in Section \ref{sec:ZN_gauge_theory} by recalling $\bZ_N$ gauge theory on the lattice, including a discussion of Wilson loops, the string tension and `t Hooft strings. We also comment on the global symmetry of these theories and their relation to $\bZ_N$ spin models, their ground-state degeneracies and the potential between static charges. 
In Section \ref{sec:phase_transitions}, we give more details on the phases of zero-temperature $\bZ_N$ gauge theories in $2+1$ dimensions, including an overview of confined and deconfined phases, and the expected nature of the transition for varying $N$.
In Section \ref{sec:NNQS}, we introduce the ideas behind neural network quantum states and compare them with traditional Euclidean Monte Carlo. Following this, we review the network architecture that we use in this work, namely lattice gauge-equivariant convolutional neural networks~\cite{Favoni:2020reg}. We describe our implementation of this and give details of sampling, initialisation, training, transfer learning and calculation of observables. Thereafter, we explain why the states represented by the networks are always in the trivial superselection sector of the $\bZ_N^{(1)}$ one-form global symmetry. With this groundwork set, in Section \ref{sec:Z2} we analyse $\bZ_2$ gauge theory. Our exploration includes computing ground-state energies as a function of coupling for varying lattice size, and expectation values of order and disorder operators and the string tension. Using finite-size scaling theory and curve collapse, we compute the critical coupling and critical exponents characterising the theory's continuous phase transition, finding excellent agreement with previous results. We then study the potential between static charges as a function of coupling, recovering the expected linear dependence on distance.
We move to $\bZ_3$ gauge theory in Section \ref{sec:Z3}. Here, we compute ground-state energies and look at order and disorder operators to identify the location of the phase transition. By looking at derivatives of the energy as a function of coupling, we argue that the transition is weakly first-order in nature. We finish with a check that the network has correctly learned charge conjugation symmetry and analyse the static charge potential. The appendices contain an overview of our implementation of Wilson loop operators, a review of calculating critical exponents from curve collapse, and a brief summary of the BST extrapolation technique which we use to identify the critical coupling.

\subsubsection{Future directions}

Obvious next steps include examining $\bZ_N$ theories for $N>3$, extending our networks to continuous gauge groups and including matter fields. In particular, there is renewed interest in $\bZ_N$ and $\Uni1$ theories under the guise of studying flux tubes using confining strings~\cite{Athenodorou:2018sab,Luo:2023cjv}, and it would be interesting to investigate if mass gaps can be computed using NNQSs. We also intend to study theories in $3+1$ dimensions; unlike approaches such as matrix product states, no conceptual changes to the network architecture are needed to move to higher dimensions.

Our identification of critical exponents for $\bZ_2$ relies on curve collapse via a leading-order finite-scaling analysis. It would be interesting to see if incorporating subleading corrections gives more precise estimates~\cite{Beach2005}, even without going to larger system sizes. In another direction, it would be encouraging to observe signs of the discontinuities that characterise a first-order transition for $\bZ_3$ by studying larger lattices. Moreover, since second Rényi entropies can be computed efficiently from a NNQS~\cite{Deng2017, Medina2021}, it would be fruitful to study the entanglement structure of the ground state and to determine the central charge of the CFTs governing the continuous phase transitions that occur for $N \neq 3$. Note that the replica method is another approach to computing the entanglement entropy \cite{Bulgarelli:2023ofi, Bulgarelli:2024onj}, and it will be insightful to compare the two techniques.

We have not fully explored how to optimise the training process. For example, we have used transfer learning when scanning over couplings. Since the form of an L-CNNs is system-size agnostic, one could also use transfer learning to move from smaller to larger lattices as demonstrated in~\cite{Luo:2021ccm, kochkov2018}. It has also been observed that for ground states with non-trivial sign structure, it is often helpful to first learn the phase of the wavefunction, with the amplitude learned towards the end of training~\cite{10.1103/PhysRevResearch.2.033075}. In a different direction, our ground-state wavefunctions were found by minimizing their energy, however one might also try to minimise the variance of the energy. Empirically, variance minimization alone seems ill-suited for finding ground states~\cite{10.1021/acs.jctc.0c00147}, though some combination of the two appears to improve performance~\cite{Zhang:2020bhg}. It may also be useful to monitor other observables such as expectation values of `t Hooft strings during training. Since the systematic error in a general observable depends linearly on the difference between the variational state and the true ground state (while the energy is quadratic in the difference), tracking other observables may provide a more sensitive measure of whether the network has converged.

By construction, L-CNNs give wavefunctions which are invariant under both gauge transformations and lattice translations. In the future, we plan to investigate whether including further symmetries improves the accuracy of the variational state. For example, the wavefunction should also be invariant under rotations and reflections, which can be implemented using group convolutional neural networks (G-CNNs)~\cite{1602.07576}. Rather than relying on the network learning this information, imposing these symmetries exactly has been found to improve results~\cite{2104.05085}. Furthermore, one can impose exact charge conjugation invariance, so that the amplitude for a lattice configuration $\mathcal{U}$ is the same as for $\mathcal{U}^\dagger$. This can be done simply by averaging the network over $\mathcal{U}$ and $\mathcal{U}^\dagger$.

\section{\texorpdfstring{$\mathbb{Z}_N$}{ZN} lattice gauge theory}\label{sec:ZN_gauge_theory}

Following \cite{Emonts:2020drm,Favoni:2020reg}, we first review pure $\bZ_{N}$ gauge theory on the lattice. In the Hamiltonian approach to lattice gauge theory, space is discretized while time remains continuous. Consider a spacelike two-dimensional lattice corresponding to a lattice gauge theory in $2+1$ dimensions, with periodic boundary conditions. The lattice spacing is taken to be one, and the size of the lattice is $L\times L$.

\begin{figure*}
\includegraphics{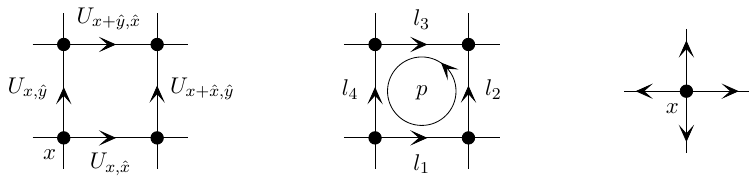}
\caption{The left figure shows our convention for associating link variables to a lattice site at $x$. The middle figure shows the links $(l_{1},l_{2},l_{3},l_{4})$ which make up a plaquette $p$ when defining the magnetic term in the Hamiltonian \eqref{eq:hamiltonian_Zn}. The right figure shows the links associated with the operator $\Theta_{x}$ acting at the lattice site at $x$, defined in \eqref{eq:gauge}, which implements Gauss's law on the lattice.}
\label{fig:links}
\end{figure*}

The gauge field degrees of freedom live on the links between the lattice sites. The link variables $U_{x,\mu}$ are valued in the gauge group. Here our notation is that $U_{x,\mu}$ determines the parallel transport from a lattice site at $x$ to a neighbouring site $x+\mu$, with $\mu=\hat{e}_{\mu}$, where the lattice spacing is taken to be one and $\hat{e}_{\mu}$ is a unit vector pointing in the positive $\mu$ direction. This is shown in Figure \ref{fig:links}. We denote a particular configuration of the $2L^{2}$ link variables by $\mathcal{U}=\{U_{x,\mu}\}$. Under a gauge transformation, the link variables transform \emph{non-locally} as
\begin{equation}
U_{x,\mu}\mapsto T_{\Omega}U_{x,\mu}=\Omega_{x}U_{x,\mu}\Omega_{x+\mu}^{\dagger},\label{eq:link_gauge_transform}
\end{equation}
where $\Omega_{x}$ are group elements.

The Hamiltonian of pure $\bZ_{N}$ gauge theory can then be written as~\cite{Horn:1979fy,Emonts:2020drm,Emonts:2022yom}
\begin{equation}
H =H_{E}+H_{B}=\frac{g^{2}}{2}\sum_{l}\left[2-(P_{l}+P_{l}^{\dagger})\right]+\frac{1}{2g^{2}}\sum_{p=(l_{1},l_{2},l_{3},l_{4})}\left[2-(Q_{l_{1}}^{\dagger}Q_{l_{2}}^{\dagger}Q_{l_{3}}Q_{l_{4}}+\text{h.c.})\right].\label{eq:hamiltonian_Zn}
\end{equation}
The first term is a sum over all links $l$ of the lattice, while the second term is a sum over all plaquettes $p$. We refer to these as the electric and magnetic terms, respectively. The links $(l_{1},l_{2},l_{3},l_{4})$ make up the plaquette $p$, as shown in Figure \ref{fig:links}. The operators $Q_{l}$ and $P_{l}$ are the standard ``clock'' and ``shift'' operators on the link $l$. These unitary operators are generalizations of the Pauli operators and satisfy a $\bZ_{N}$ algebra:
\begin{equation}
P_{l}^{\dagger}P_{l}=Q_{l}^{\dagger}Q_{l}=\id,\qquad P_{l}^{N}=Q_{l}^{N}=\id,\qquad P_{l}^{\dagger}Q_{l}P_{l}=\ee^{-2\pi\ii/N}Q_{l},
\label{eq:ZN_ops}
\end{equation}
where operators that act on different links commute with one another.

The discretised version of Gauss's law for a $\bZ_{N}$ gauge theory is encoded by a set of local unitary vertex operators $\Theta_{x}$. For $d=2$, these operators are defined as
\begin{equation}
\Theta_{x}=P_{x,\mu}P_{x,\nu}P_{x-\mu,\mu}^{\dagger}P_{x-\nu,\nu}^{\dagger},\label{eq:gauge}
\end{equation}
where $\mu=\hat{x}$ and $\nu=\hat{y}$, so that the links correspond to those in Figure \ref{fig:links}. The lattice Hamiltonian commutes with all of these operators,
\begin{equation}
[\Theta_{x},H]=0\quad\forall\:x,
\end{equation}
which implies the local gauge invariance of the theory. The vertex operators have eigenvalues $\ee^{2\pi\ii n/N}$, with operators for different sites commuting. For a state $|\Psi\rangle$, Gauss's law is the statement that
\begin{equation}
\Theta_{x}|\Psi\rangle=|\Psi\rangle\quad\forall\,x.\label{eq:gauss}
\end{equation}
A state which satisfies this constraint is gauge invariant. The ground state of the theory is always gauge invariant. There are, however, other superselection sectors of the Hilbert space of the theory, classified by the eigenvalues of $\Theta_{x}$. For example, the twisted or charged sector where $\Theta_{x}=\ee^{2\pi\ii n/N}$ and $\Theta_{x'}=\ee^{2\pi\ii n'/N}$ for integer $n$ and $n'$ can be interpreted as the gauge-invariant states with charges $n$ and $n'$ located at the lattice sites $x$ and $x'$ respectively.

The link variables $U_{x,\mu}$ are elements of $\bZ_{N}$, and so can be thought of simply as phases of the form $\ee^{2\pi\ii q/N}$, where $q\in\{0,\dots,N-1\}$. With this, we define states $|q\rangle_{l}$ which span the Hilbert space on the link $l$. These states are assumed to give an orthonormal basis for the Hilbert space, $\langle q|q'\rangle=\delta_{q,q'}$ and are eigenstates of the clock operator:
\begin{equation}
Q_{l}|q\rangle_{l}=\ee^{2\pi\ii q_{l}/N}|q\rangle_{l}.
\end{equation}
Moreover, the shift operator $P$ acts as a periodic lowering operator, 
\begin{equation}
P_{l}|q\rangle_{l}=|q-1\rangle_{l},
\end{equation}
with $P_{l}|0\rangle_{l}=|N-1\rangle_{l}$. We also define plaquette variables $U_{x,\mu\nu}$ as $1\times1$ untraced Wilson loops of the form
\begin{equation}
U_{x,\mu\nu}=U_{x,\mu}U_{x+\mu,\nu}U_{x+\nu,\mu}^{\dagger}U_{x,\nu}^{\dagger}.
\end{equation}
Under \eqref{eq:link_gauge_transform}, the plaquette variables transform \emph{locally} as
\begin{equation}
U_{x,\mu\nu}\mapsto T_{\Omega}U_{x,\mu\nu}=\Omega_{x}U_{x,\mu\nu}\Omega_{x}^{\dagger}.\label{eq:plaqs}
\end{equation}
For our example of a $d=2$ lattice, we take $\mu=\hat{x}$ and $\nu=\hat{y}$ so that the loop is traversed anticlockwise, in agreement with the convention in Figure \ref{fig:links}.

At a given time, a wavefunction $\Psi$ for the lattice system takes in a configuration $\mathcal{U}\in\mathbb{Z}_{N}^{2L^{2}}\subset\bC^{2L^2}$ of the link variables and returns a complex number:
\begin{equation}
\Psi\colon\bC^{2L^{2}}  \rightarrow \mathbb{C},\quad\mathcal{U} \mapsto\Psi(\mathcal{U}).
\end{equation}
We will be particularly interested in those wavefunctions which can approximate the ground state of the system. Since the Hamiltonian is time independent, the ground state is time independent, so it is sufficient to consider only the two-dimensional spatial lattice for a $2+1$-dimensional theory. The ground state must also be gauge invariant. Denoting the ground state by $|\Psi_0\rangle$, this means that under the gauge transformation \eqref{eq:link_gauge_transform}, the wavefunction $\Psi_0$ is invariant, i.e.~$\Psi_0(T_{\Omega}\mathcal{U})=\Psi_0(\mathcal{U})$.\footnote{More generally, a gauge equivariant function $f$ obeys $f(T_{\Omega}\mathcal{U})=T_{\Omega}f(\mathcal{U})$.}

\begin{figure}
\includegraphics{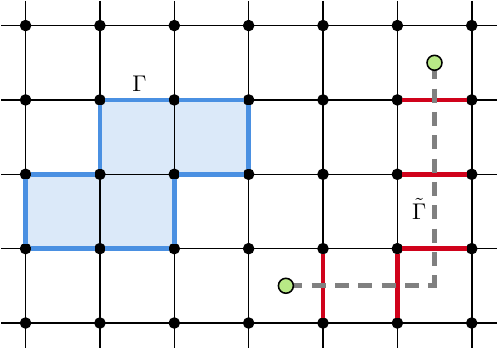}

\caption{Order and disorder operators on the lattice. The blue path denotes a Wilson loop, $W_\Gamma = \prod_{\Gamma} Q$, given by the product of clock operators along the closed contour $\Gamma$. The shaded blue region shows the area enclosed by the Wilson loop. The dotted gray path denotes a `t~Hooft string, $T_{\tilde\Gamma}=\prod_{\tilde{\Gamma}}P$, given by the product of shift operators on the links pierced by the contour $\tilde\Gamma$ between two points on the dual lattice. }
\label{fig:order_disorder_lattice}
\end{figure}

\subsection{Wilson loop operators}\label{sec:wilson}

Naively, one might expect that, in analogy to the magnetization of a spin system, the phases of a lattice gauge theory should be distinguished by a spontaneous alignment of the link degrees of freedom. This would manifest as a non-zero value of $\sum_l\langle Q_l\rangle$. In the analogous spin system, a non-zero magnetization indicates a spontaneous breaking of a global symmetry. In the gauge theory, a non-zero expectation value would instead imply the spontaneous breaking of a local symmetry. This is impossible by Elitzur's theorem~\cite{Elitzur:1975im}, and so the phases of the theory cannot be distinguished by these expectation values (which vanish by gauge invariance). Instead, one must look at non-local, gauge-invariant observables, the simplest of which is the Wilson loop.

A Wilson loop operator $W_{\Gamma}$ is defined given $\Gamma$, a closed oriented path on the lattice~\cite{Wilson:1974sk}. Explicitly, we define
\begin{equation}
W_{\Gamma}=\prod_{l\in\Gamma}Q_{l},
\end{equation}
where our convention is that if the link $l$ is oriented to agree with the arrows in Figure \ref{fig:links}, one has $Q_{l}$, whereas if the link has the opposite orientation, one instead takes the Hermitian conjugate, $Q_{l}^{\dagger}$. Note that the magnetic term $H_B$ in the Hamiltonian is simply $(1-\re W_\square)$, summed over the $1\times1$ plaquettes $\square$. An example of a Wilson loop on the lattice is shown in Figure \ref{fig:order_disorder_lattice}.

In a pure $\bZ_{N}$ gauge theory, the expectation values of Wilson loop operators can detect the phase of the system~\cite{Wilson:1974sk,Fradkin:1978th, Ritz-Zwilling:2020itc}. In particular, whether the system is in a confined or deconfined phase is determined by the scaling of the expectation value of $W_\Gamma$ as the size of the loop $\Gamma$ is varied. Therefore, the Wilson loop serves as an (unconventional) order operator for the lattice gauge theory. In a deconfined phase, the magnetic term $H_B$ in the Hamiltonian dominates and the expectation value of the Wilson loop decays exponentially with the length of the perimeter $P_{\Gamma}$ of the loop as
\begin{equation}\label{eq:perimeter_law}
\langle W_{\Gamma}\rangle\sim\ee^{-\kappa P_{\Gamma}},
\end{equation}
where $\kappa$ is coupling dependent. In a confined phase, the electric term $H_E$ dominates, leading to an additional confining area-law scaling:
\begin{equation}\label{eq:area_law}
\langle W_{\Gamma}\rangle\sim\ee^{-\kappa P_{\Gamma}-\sigma A_\Gamma},
\end{equation}
where $A_\Gamma$ is the area of the loop and $\sigma$ is known as the string tension. Again, $\sigma$ is expected to depend on the coupling $g$. Note that the Wilson loop does not follow a strict area law in the confined phase since there is always a perimeter law contribution.

\subsection{Creutz ratio}

In the thermodynamic limit, one expects an order parameter to be positive in an ordered phase and zero in a disordered phase. The expectation values of Wilson loop operators do not display this behaviour, and their interpretation is more nuanced. Instead, as we discussed above, it is the scaling with the size of the corresponding loop that is different in each phase. However, this behaviour can be used to extract the string tension, which gives a conventional order parameter. It is simple to check that the so-called Creutz ratio \cite{Creutz:1980wj}
\begin{equation}
    \chi_{l}=-\ln\frac{\langle W_{l\times l}\rangle\langle W_{(l-1)\times(l-1)}\rangle}{\langle W_{(l-1)\times l}\rangle\langle W_{l\times(l-1)}\rangle},
\end{equation}
where $W_{l\times m}$ is an $l\times m$ Wilson loop operator on the lattice, removes the perimeter scaling and so computes the string tension, $\chi_{l}=\sigma$. Due to this, the Creutz ratio behaves like a standard order parameter in the thermodynamic limit: positive in the confining phase and vanishing in the deconfined phase~\cite{Gonzalez-Arroyo:2012euf}. Note that the perimeter- and area-law scalings given in \eqref{eq:perimeter_law} and \eqref{eq:area_law} are only the leading contributions. The Creutz ratio is constructed to remove some subleading finite-size corrections and also corrections that come from edge/corner effects. The leftover corrections are smaller for larger Wilson loops, so that $\chi_l$ is a better estimate of the continuum string tension for larger values of $l$. 

Unfortunately, computing $\chi_l$ using Monte Carlo sampling can be challenging. In the confined phase, where $\chi_l$ is expected to be non-zero, Wilson loops decay both with the area of the loop and increasing coupling. The string tension is then estimated by computing a ratio of very small numbers, which can be very sensitive to Monte Carlo errors -- we will see this later in Figure \ref{fig:Z2_creutz}. Consequently, one often restricts the computation to smaller values of $l$.

\subsection{`t~Hooft operators}

A string of $P$ operators, $T=\prod_{\tilde{\Gamma}}P$, along an open path $\tilde{\Gamma}$ between two points on the \emph{dual} lattice defines a magnetic line operator known as a `t~Hooft string. These `t~Hooft strings serve as disorder operators for the gauge theory, and can diagnose the confined-deconfined phases. An example of a `t~Hooft string operator is shown in Figure \ref{fig:order_disorder_lattice}.

The insertion of a Gauss's law operator $\Theta_x$ acts on a `t~Hooft string simply by shifting the path $\tilde\Gamma$ without moving its endpoints. Furthermore, since the ground state is gauge invariant, insertions of $\Theta_x$ are ``free'' and do not change expectation values. Thus, $\langle T_{\tilde\Gamma} \rangle$ depends only the endpoints of the string and not on the path itself. This can be interpreted as the string operator creating a pair of quasi-particles (magnetic monopoles) which reside on the plaquettes at either end of the string, corresponding to the green dots in the example in Figure \ref{fig:order_disorder_lattice}. Note that the distance between the end points must scale with the lattice size $L$ in order to obtain a line operator in the limit as $L \to \infty$. In a deconfined phase, the expectation value of a `t~Hooft string decays exponentially with the distance between its endpoints. In a confined phase, $\langle T_{\tilde\Gamma} \rangle$ is independent of distance, since the monopoles living at the ends of the `t~Hooft string are condensed.

\subsection{Global symmetry and dual spin models}\label{sec:spin_models}
$\bZ_{N}$ gauge theory in $2+1$d can be obtained by gauging the $\bZ_{N}^{(0)}$ zero-form global symmetry of a $\bZ_{N}$ spin model. The $\bZ_{N}^{(0)}$ symmetry corresponds to transforming all spins of the model by a global $\bZ_{N}$ factor. In the case of the Ising model, we have a $\bZ_{2}$ symmetry that corresponds to flipping all the spins. By Poincaré duality, we expect that the theory after gauging will have a dual $\bZ_{N}^{(1)}$ one-form global symmetry in $2+1$ dimensions~\cite{Gaiotto:2014kfa}. The phase transitions that occur in $\bZ_{N}$ gauge theories are characterised by breaking or restoration of this one-form global symmetry. In the following subsection, we will use this one-form global symmetry to study the ground-state degeneracy of these theories at small coupling.

On a torus, these global symmetry operators are magnetic `t~Hooft \emph{loops} wrapping non-contractible cycles. Similar to `t~Hooft strings, these operators are topological and depend on only the homotopy class of the loops. However, unlike the open string operators, the closed loops intersect each plaquette exactly twice and, as a result, they commute with the magnetic term of the Hamiltonian. Since they are built from the $P$ operators, they manifestly commute with the electric term of the Hamiltonian and thus commute with the entire Hamiltonian for all values of the coupling $g$~\cite{Sachdev:2018ddg, Rayhaun:2021ocs}. 

As a result of the gauging procedure, the \emph{local} operators that transform non-trivially under the $\bZ_{N}^{(0)}$ zero-form global symmetry are projected out.  In contrast, extended line operators with these charges at their end points are a part of the spectrum of the gauge theory \cite{Chang:2018iay}. Examples of such extended operators include the Wilson and `t~Hooft line operators discussed earlier. Since local gauge symmetries cannot break spontaneously on a finite lattice, these extended operators can serve as order parameters for diagnosing phase transitions.  

\subsection{Ground-state degeneracy}\label{sec:degeneracy}

\begin{figure}
\includegraphics{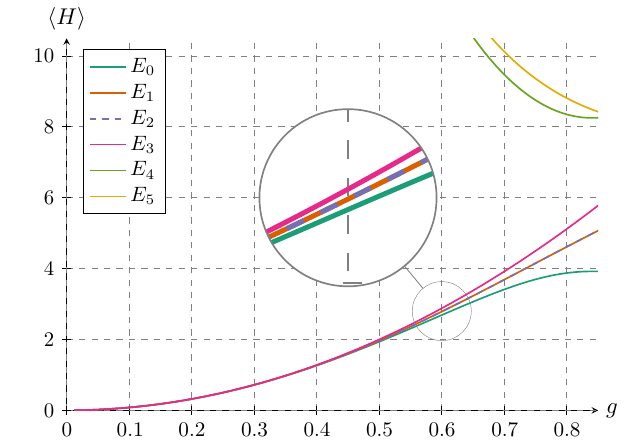}

\caption{Low-lying energy levels of $\bZ_2$ lattice gauge theory computed using exact diagonalization on an $L=2$ lattice. Away from $g=0$, there is a unique ground state (teal), with an approximate fourfold degeneracy for small but non-zero couplings.}
\label{fig:energy_Z2_2d_degen}
\end{figure}

$\bZ_{N}$ gauge theory has interesting low-lying state structure on lattices with non-trivial topology, such as the torus defined by our periodic boundary conditions. Focusing on $\bZ_2$ for concreteness, these states can be discerned from the global symmetry operators wrapping the two non-contractible cycles of the torus~\cite{Sachdev:2018ddg}. The magnetic `t~Hooft loops which wrap the $x$ and $y$ directions of the torus are denoted by $V_{x}$ and $V_{y}$ respectively. As with `t~Hooft strings, these operators are topological and depend on only the homotopy class of the loops. Since these two operators commute with the Hamiltonian, even when the eigenstates of $H$ are degenerate one can find a basis of states which are simultaneously eigenstates of the Hamiltonian and $V_{x,y}$. 

In the $g\to\infty$ limit, the ground state $|\Psi_0\rangle$ is given by all links in the $+1$ eigenstate of the shift operator ($X$ for $\bZ_2$), and so the ground state has eigenvalue $+1$ for both $V_{x}$ and $V_{y}$. States with different $V_{x,y}$ eigenvalues must have at least one spin in the $-1$ eigenstate of the $X$ operator, which will cost an energy $\sim g^{2}$. As a result, these states cannot be degenerate with the ground state, so that $|\Psi_0\rangle$ is unique even in the $L\to\infty$ limit.

As $g\to0$, we can express the ground state in terms of the basis of the clock operator which is $Z$ for $\bZ_2$. By taking linear combinations, we can express the four lowest energy states as $(1 \pm V_x)(1 \pm V_y) |0\rangle^{\otimes 2L^2}$ with eigenvalues $V_x = \pm 1$ and $V_y = \pm 1$. At $g=0$, these states have the same energy, giving a fourfold degeneracy for the ground state. For small but non-zero couplings, these $V_{x,y}$ eigenstates are no longer degenerate, but separated by a splitting that scales as $g^{4L}$ with the size $L$ of the lattice~\cite{Sachdev:2018ddg}. As a result, the splitting becomes exponentially small deep in the deconfined phase, $g \ll 1$. This effect occurs since there is non-zero tunneling amplitude between states with distinct values of $\mathbb{Z}_{2}$ flux. In the general case of $\mathbb{Z}_{N}$ gauge theory on a closed compact oriented surface $\Sigma$ with first homology group $H_{1}(\Sigma, \mathbb{Z}_{N})$, the ground state degeneracy for $g=0$ is given by~\cite{Bullock:2006bv}
\begin{equation}
    \text{Ground-state degeneracy on }\Sigma  = N^{|H_{1}(\Sigma, \mathbb{Z}_{N})|},
\end{equation}
where $|H_{1}(\Sigma, \mathbb{Z}_{N})|=2$ if $\Sigma$ is a torus. This degeneracy is one of the characteristic features of $\bZ_{N}$ topological order~\cite{Read:1991zz,Wen:1991sca,Bais:1991pe,Maldacena:2001ss,Kitaev:1997wr,Freedman:2004lda,Hansson:2004wca}.

The behaviours we have discussed actually persist throughout the deconfined and confined phases. In general, for $g \neq 0$ the ground state is unique and has eigenvalue $+1$ for both $V_{x}$ and $V_{y}$. Below the critical coupling, for $\bZ_2$, there is an approximate fourfold degeneracy with a splitting governed by the size of the lattice that goes to zero as $g\to0$. This can be seen in Figure \ref{fig:energy_Z2_2d_degen}, which shows the lowest-lying states of  $\bZ_2$ gauge theory on a $2\times2$ lattice computed using exact diagonalization. One observes the approximate fourfold degeneracy at small couplings, with the gap growing with $g$.

\subsection{Potential energy between charges}\label{sec:potetial}

As reviewed earlier, the uncharged sector of the Hilbert space, describing the gauge field degrees of freedom, is the subsector which satisfies $\Theta_{x}|\Psi\rangle=|\Psi\rangle$ for all lattice sites $x$.  We now discuss the subsectors with static charges placed at the lattice sites. Since the insertion of a single charge at a lattice site is not gauge invariant, the simplest setup one can consider is an open Wilson line with opposite charges at each end. Starting from a gauge-invariant state $|\Psi\rangle$, one can introduce charges by acting with a string of clock operators that stretch between lattice sites on which the charges are placed. For example, the following string operator can be used to 
place the charges a distance $r$ apart along the $x$-direction:
\begin{equation}\label{eq:wilson_string}
W_{r}=Q_{x,\mu}Q_{x+\hat{x},\mu}\dots Q_{x+r\hat{x},\mu}.
\end{equation}
Note that the charges in the $\mathbb{Z}_{N}$ theory are constrained to be $\{0, 1,\ldots, N-1\}$. Using the $\bZ_{N}$ commutation relations, one finds
\begin{equation}
\Theta_{x}(W_{r}|\Psi\rangle)=\ee^{-2\pi\ii/N}(W_{r}|\Psi\rangle),\qquad
\Theta_{x+r\hat{x}}(W_{r}|\Psi\rangle)=\ee^{2\pi\ii/N}(W_{r}|\Psi\rangle).
\end{equation}
From these phases, one deduces that the Wilson line connects static charges at $x$ and $x+r\hat{x}$, with charges $N-1$ and $1$ respectively. To compute the potential between the charges, one can compare the ground-state energies with and without the charges. In practice, this means one compares
\begin{equation}
\langle H\rangle_0\quad\text{vs}\quad\langle W_{r}^{\dagger}HW_{r}\rangle_{0'},
\end{equation}
where the expectation values are evaluated using the ground-state wavefunction in each sector, denoted by $0$ and $0'$ respectively.

Let us now analyse the potential energy $V(r)$ needed to separate the charges in the limit of very small and large values of $g$. When $g \ll 1$, the magnetic term in the Hamiltonian \eqref{eq:hamiltonian_Zn} dominates and as a result the ground state is an eigenvector of $Q_{l}$ with eigenvalue $+1$ for all links $l$. Consequently, the Wilson line operator $W_{r}$ leaves the ground state unchanged. Therefore, pulling the charges out of the vacuum and separating them does not take any energy, thus the charges are deconfined. 

In the opposite limit, when $g \gg 1$, the electric term in the Hamiltonian \eqref{eq:hamiltonian_Zn} dominates and the ground state is an eigenvector of $P_{l}$ with eigenvalue $+1$ for all links $l$. Employing the commutation relations \eqref{eq:ZN_ops}, we observe that the $Q$ operators forming the Wilson line \eqref{eq:wilson_string} modify the eigenvalues of $P_{l}$ and $P_{l}^{\dagger}$ along the Wilson line to $\ee^{2\pi\ii/N}$ and $\ee^{-2\pi\ii/N}$, respectively. To first-order in perturbation theory, the ground-state energy in the twisted sector is given by evaluating $W_{r}^{\dagger}HW_{r}$ using the untwisted ground-state wavefunction.\footnote{Experimentally, we have observed negligible difference between evaluating the expectation of $W^\dagger_r H W_r$ using the ``true'' ground state in the twisted sector and the ground state of the sector without static charges, at least for couplings in the range $0.5$ to $1.1$. Using the $\bZ_N$ operator identities, the result of conjugating the Hamiltonian by the Wilson string operator is $W^\dagger_r H W_r = H + \frac{g^2}{2}\sum_{i=0}^{r-1} \bigl(P_{x+i,\mu}(1-\ee^{-2\pi\ii/N}) + \text{h.c.}\bigr)$, where the second term can be thought of as a perturbation. Since the original Hamiltonian is given by a sum over $2L^2$ links, while the perturbation is a sum over only $r$ links, the parameter controlling the size of the perturbation is $r/2L^2$ which will be small for large enough lattices. Providing the gap between the original ground state and the first excited state is large compared with the energy shift due to the perturbation, one will get a good estimate of the perturbed ground-state energy using first-order perturbation theory, where one computes the correction by evaluating the expectation value of the perturbation using the unperturbed ground state.} Plugging this back into the Hamiltonian, and neglecting the magnetic term which is very small in this limit, one finds that the energy difference is linear in the distance $r$
\begin{equation}\label{eq:potential_energy} 
    V(r) \approx \langle W_{r}^{\dagger}HW_{r}\rangle_0 - \langle H\rangle_0 = g^2 r \bigl(1 - \cos(2\pi/N)\bigr). 
\end{equation}
We see that it costs a great deal of energy to increase the separation between the charges, implying that the charges are confined. Note that one can instead place charges $k$ and $N-k$ on the two ends by acting with the operator $W_{r}^{k}$. In this case, the potential energy in the confined phase is given by $V(r)\approx g^2 r (1 - \cos(2\pi k/N))$.

In summary, the energy needed to separate the charges differs in the two phases. In the confined phase, the potential energy rises linearly with the distance between the charges, corresponding to the formation of a confining flux tube or string between the charges. In the deconfined phase, the electric flux lines are condensed and the potential between the charges is very small, becoming independent of distance in the limit of large $L$.

It is useful to compare this perspective with Euclidean lattice gauge theory, where a timelike Polyakov loop (a Wilson loop extended in imaginary time) serves as an order parameter for confinement. The expectation value of this non-local operator is then related to the free energy of an isolated static charge. In particular, it should be zero in the confined phase, indicating an infinite free energy cost for isolating a single charge. Conversely, a non-zero Polyakov loop expectation value signals deconfinement, where isolated charges can exist. To directly study confinement, one instead considers the correlation function of two timelike Polyakov loops separated in space. If the theory is in the confined phase, this correlation function should decay exponentially with the separation, reflecting the linear confining potential between the static charges represented by the Polyakov loops. Note that it is only in the continuum limit where Lorentz invariance emerges that one can directly relate spatial and timelike Wilson loops, and thus identify the string tension $\sigma$ appearing in \eqref{eq:area_law} with the coefficient of the linear potential between static charges. Away from this limit, these two quantities do not necessarily agree.

\section{Phases and phase transitions in \texorpdfstring{$\mathbb{Z}_{N}$}{ZN} theories }\label{sec:phase_transitions}

The central aim of this paper is to use neural network quantum states to model the ground-state wavefunction of $\bZ_N$ lattice gauge theories. Beyond simply computing the energy of the ground state, one can use this approach to study the phases and phase transitions as the coupling $g$ is varied. With this in mind, we now review some useful concepts for studying transitions and critical behaviour on the lattice.

\subsection{First-order vs continuous phase transitions}

In quantum mechanics, a first-order phase transition is characterised by level-crossing in the energy spectrum. This means that, as a parameter in the Hamiltonian (such as the coupling $g$) is varied, the energy levels of two distinct states intersect. At the point of intersection, the ground state of the system changes abruptly from one state to another. The ground-state energy of the system
\begin{equation}
    E_0 = \langle H \rangle_0 \equiv \frac{\langle \Psi_0 | H | \Psi_0\rangle}{\langle \Psi_0 | \Psi_0\rangle},
\end{equation}
has a kink at the transition point, while the derivative of the energy with respect to the coupling is discontinuous. This abrupt change is the hallmark of a first-order phase transition. For the case of the first-order transitions, the energy gap between the lowest energy state and the first excited state remains finite even as the ground state energy levels cross and as a result of this gap the system has finite correlation length. 

For a lattice gauge theory, in the limit where the lattice size is much greater than the correlation length of the system, one expects a sharp change between an ordered and a disordered state.  Moreover, the order parameters which characterise the order and disordered states also display a discontinuity at the transition. For finite $L$, the ordered and disordered states coexist in the vicinity of the transition. The sharp kink in the energy is smoothed somewhat, with the derivative of the energy displaying a large gradient at the transition point. 

Unlike a first-order transition, a continuous quantum phase transition occurs without level-crossing. Here, the ground state of the system evolves smoothly as a function of the coupling. There is no abrupt change in the ground state, but rather a gradual change in its properties. The smooth change of the state is mirrored in the smooth change of the ground-state energy, in contrast to the kink/discontinuity at a first-order transition. When such a continuous phase transition occurs, the spectral gap between the ground state and the first excited state closes. Consequently, the correlation length of the system diverges, leading to the onset of long-range correlations across the system \cite{Sachdev2011}. Due to the resulting scale invariance of the system, continuous critical phenomenon can be characterised using critical exponents, which capture the behaviour of physical quantities near the phase transition. For many quantum-mechanical lattice models, such as lattice gauge theory, the system at the critical coupling has an enhanced conformal symmetry in addition to scale invariance leading to a conformal field theory (CFT)~\cite{ZinnJustin2002}. 

\begin{figure}
    \includegraphics{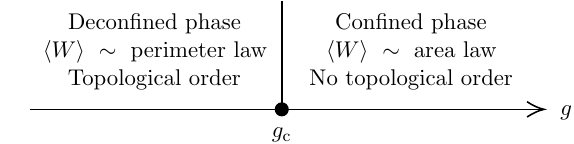}
    \caption{Phases of $\mathbb{Z}_{N}$ gauge theory in $2 + 1$ dimensions. The deconfined and confined phases are separated by a confinement transition as the critical coupling $g_{c}$.}
    \label{fig:phases}
\end{figure}

\subsection{Confined and deconfined phases}\label{sec:phases}

In $2 + 1$ dimensions, discrete gauge theories with a Hamiltonian given by \eqref{eq:hamiltonian_Zn} have two possible phases. They are in a deconfined phase at weak coupling and in a confined phase at strong coupling. As explained in Section \ref{sec:potetial}, in the deconfined phase it takes very little energy to separate the charges, while in the confined phase the energy required to pull charges apart grows linearly with the length of the flux string. The expectation values of Wilson loops follow a perimeter law in the deconfined phase and an area law in the confined phase, as described in Section \ref{sec:wilson}. Since the magnetic monopoles are in condensed in the deconfined phase, the value of a `t Hooft string operator is independent of distance in this phase. The deconfined phase is topologically ordered, with a ground state degeneracy that depends on the topology of the system. This topologically ordered state has long-range quantum entanglement \cite{Chen2010}. The $\mathbb{Z}_{N}^{(1)}$ one-form symmetry is preserved in the confined phase, while it is broken in the deconfined phase. These characteristics of the two phases are summarised in Figure \ref{fig:phases}.

\subsection{Confinement transition}\label{sec:crit_behaviour}

The nature of the confinement phase transition depends on $N$, with a first-order transition for $N=3$ and a continuous transition occurring for all other values of $N$~\cite{Bhanot:1980, Omero:1982, Borisenko:2013sqa}. Evidence for this comes from Monte Carlo simulation of spin systems, which are dual to these lattice gauge theories. Furthermore, high-order perturbative expansions confirm the Monte Carlo simulation results~\cite{Vidal:2008uy, Schulz:2011tk}. As explained in Section \ref{sec:spin_models}, this duality corresponds to gauging the global $\mathbb{Z}_N$ zero-form symmetry in the spin-system to obtain the lattice gauge theory. As a result, in the case of a continuous transition the theory at the critical coupling $\gc$ is described by the gauged version of the CFT that describes the continuous transition in the corresponding spin system.  

The second-order phase transition for the transverse-field Ising model is in the universality class of $2+1$-d Ising Wilson--Fisher CFT. Thus, the confinement phase transition for the $\mathbb{Z}_2$ lattice gauge theory belongs to the $2+1$-d gauged Ising universality class \cite{Sachdev:2018ddg}. This theory is also sometimes referred to as the Ising$^{\ast}$ theory. In Section \ref{sec:Z2}, we calculate the critical exponents for this transition using neural network quantum states. For $\mathbb{Z}_3$, we observe the first-order nature of the phase transition and determine the approximate location of the critical coupling in Section \ref{sec:Z3}. 

Remarkably, the $\mathbb{Z}_4$ theory decomposes into two decoupled $\mathbb{Z}_2$ theories for every value of the coupling, with the critical point hosting two copies of the $2+1$-d Ising$^{\ast}$ CFT~\cite{Grosse:1980fg}. Monte Carlo simulations suggest that the continuous transitions for $N>4$ belong to the $2+1$-d XY ($\Orth2$ model) universality class~\cite{Campostrini:2000iw, Borisenko:2013sqa}. The critical coupling at which the confinement transition occurs goes to zero as the value of $N$ increases. The $N \to \infty$ limit recovers $\Uni1$ lattice gauge theory, which is confined for all values of the coupling~\cite{Bhanot:1980}. As shown by Polyakov, this is a result of the fact that certain monopole operators are relevant, leading to a phase of unbroken $\bZ_{N}^{(1)}$ one-form global symmetry \cite{Polyakov:1975rs, Polyakov:1976fu}. The phase diagram is much richer in $3 + 1$ dimensions, with two phase transitions occurring for $N \geq 5$ \cite{Creutz:1979zg, Apte:2022xtu}. Investigating these phase transitions in $3 + 1$ dimensions with neural network quantum states will be a fruitful direction for future work. 

Note that the critical couplings (or inverse temperatures) at which the transition occurs computed with Monte Carlo methods are not directly comparable to the critical couplings that one finds from a Hamiltonian approach. The reason for this is that the time direction of the $d+1$-dimensional spacetime is also discretized. In order to recover the Hamiltonian perspective, one must take a certain anisotropic limit of the lattice~\cite{PhysRevE.66.066110,Tupitsyn:2008ah}, which comes with a non-trivial rescaling of the couplings of the theory. For this reason, other than for simple examples, it is difficult to directly compare the location of phase transitions in the two approaches.

\section{Neural network quantum states}\label{sec:NNQS}
\begin{table*}[htb]
\centering
\begin{tabular}{cc}
\hline
Monte Carlo & Neural network quantum states \\
\hline
Computes path integral & Computes ground state \\  
$d+1$-dimensional spacetime lattice & $d$-dimensional spatial lattice \\ 
Sign problem & No sign problem \\
Asymptotic convergence guarantee & No convergence guarantee \\ 
Difficult sampling problem & Difficult optimisation problem \\
Few hyperparameters & Many hyperparameters \\
\hline
\end{tabular}
\caption{Brief comparison of path-integral Monte Carlo (MC) and neural network quantum state (NNQS) approaches.}
\label{tab:mc_vs_nnqs}
\end{table*}

There is a long history of using the variational method to find approximations to the ground-state wavefunction of quantum-mechanical systems~\cite{Tilly:2021jem}. This relies on the observation that, given the system's Hamiltonian $H$ and a normalisable wavefunction $\Psi_\theta$, where $\theta$ are parameters that specify the particular wavefunction from a family of variational states, the functional\footnote{Here we are explicitly dividing by the norm of the variational state. For ease of notation, we often suppress this and assume that the state is correctly normalized, though in practice it is simpler to work with unnormalized NNQSs.}
\begin{equation}
    E_\theta = \langle H \rangle_\theta \equiv \frac{\langle \Psi_\theta | H | \Psi_\theta \rangle}{\langle \Psi_\theta |\Psi_\theta \rangle},
\end{equation}
is bounded from below by the true ground-state energy $E_0$ of the system, with $E_\theta=E_0$ if and only if $\Psi_\theta$ is the exact ground-state wavefunction of the system. Thus, by minimizing $E_\theta$ with respect to the parameters $\theta$, one obtains an approximation to both the ground-state energy and wavefunction itself. 

The more flexible the variational ansatz, the better the resulting approximation should be. Moreover, if the form of $\Psi_\theta$ is particularly well suited to the system, one expects that fewer parameters are needed to obtain a good approximation. For example, for the simple harmonic oscillator, an ansatz of a Gaussian in the square of the displacement needs only a single parameter to describe the exact ground state, while an ansatz in terms of Fourier modes would need more parameters (and an infinite number to capture the exact ground state). Finding a good approximation to the ground state of a complicated quantum system then requires variational states which are well suited for the system under consideration and flexible enough to capture the relevant physics.

Neural network quantum states (NNQS) were developed in 2016 by Carleo and Troyer as a new kind of variational ansatz for quantum systems~\cite{1606.02318}. Their idea was to use a neural network as the trial wavefunction of a quantum system, with the parameters of the network chosen to minimise the expectation value of the Hamiltonian. In their most basic form, neural networks give a map from inputs to outputs, with the map given by compositions of linear transformations and non-linear activation functions. In our case, we are interested in networks which map from a configuration of link variables to a single complex number, $\bC^{2L^2}\to\bC$, so that the network can be interpreted as assigning a probability amplitude to a given lattice configuration. The linear transformations can be thought of as acting with matrices whose entries are known as ``weights'', which are then interpreted as variational parameters. The non-linear activation functions, often alternated with the linear transformations, result in the network output being a complicated non-linear function of the variational parameters. Thanks to this, NNQSs can capture a wide range of phenomena, including ground-state wavefunctions.

A NNQS  gives a variational ansatz $\Psi_\theta$ for the wavefunction parametrised by the choice of weights $\theta$. The next question is how to choose these weights so that $\Psi_\theta$ provides an approximation to the ground state of a given quantum system. Again, one can use the variational method by trying to minimise the expectation value of the Hamiltonian with respect to the weights of the neural network. Since the network is a complicated non-linear function of these weights, and these weights often number in the thousands or tens of thousands, it is not possible to solve this minimization problem exactly. Instead, one resorts to numerical methods to iteratively reduce $E_\theta$ by varying the weights. The way to do this is (stochastic) gradient descent. The key to this is the fact that modern neural network packages allow for automatic differentiation, so that one can differentiate $E_\theta$ with respect to the weights $\theta$ and evaluate the resulting gradient exactly (that is, without using finite differences). Given this gradient, one adjusts the weights to move in the direction of steepest descent. By repeatedly iterating, one hopes to ``train'' the network and eventually find the set of weights which minimise $E_\theta$. At the end of training, one not only has an estimate of the ground-state energy, but also an approximation $\Psi_\theta$ for the exact ground-state wavefunction $\Psi_0$.

\subsubsection{MC vs NNQS}

Before describing the gauge-invariance networks we will use in this paper, let us quickly compare traditional Monte Carlo (MC) methods with neural network quantum states (NNQS) for studying lattice theories:
\begin{itemize}
    \item Dimensionality of the lattice: In NNQS, the trial wavefunction is defined on the $d$-dimensional spatial lattice. In MC, fields are defined on a $d+1$-dimensional lattice that includes the imaginary time direction.
    \item Variational principle: NNQS relies on a variational principle, wherein the trial wavefunction is optimised to minimise the energy expectation value. MC does not involve any optimisation, but instead aims to directly evaluate the path integral on the lattice using Monte Carlo sampling.
    \item Sampling problem: In NNQS, one generates configurations of gauge fields on a $d$-dimensional lattice weighted according to the trial wavefunction. In MC, one instead samples $d+1$-dimensional lattice configurations weighted according to the Euclidean action $\ee^{-S}$. 
    \item Sign problem: In Monte Carlo (MC) simulations, a sign problem occurs because the fermion determinants in the path integral can turn complex in the presence of a finite chemical potential or a non-zero theta angle. This complexity makes it difficult to treat $\ee^{-S}$ as a straightforward probability measure, which prevents the use of standard Monte Carlo methods. Moreover, even when one can absorb the complex phase, evaluating expectation values often requires extremely high accuracies due to possible cancellations. This sign problem is absent in NNQS, since one samples from the Born distribution $|\Psi|^2$, meaning that standard Monte Carlo sampling can be used.\footnote{Note that one can still run into sign problems when considering operators with non-trivial phase structure, so that expectation values rely on many cancellations, leading to large variances in Monte Carlo estimates.}
    \item Convergence: In the absence of sign problem, MC is guaranteed to converge asymptotically when a sufficient number of samples have been drawn. Nevertheless, number of steps required to adequately sample the system increases as the critical point is approached due to critical slowing, making simulations near criticality computationally expensive \cite{Wolff:1989wq}. In contrast, a fixed network has limited expressivity \cite{Guhring2022} and may not capable of representing the true ground state of a system. Furthermore, even if a network is capable of representing the ground state, there is no guarantee that gradient based optimisation methods will efficiently find the optimal set of parameters.

    \item Systematic improvements: In MC, systematic improvements come from improving the Monte Carlo sampling and summing over more lattice configurations. The accuracy of NNQS can be systematically improved by using more sophisticated trial wavefunctions with more variational expressivity. However, there is a trade-off between expressivity and the efficiency of optimisation. 
    
    \item Observables: In MC, observables are calculated as ensemble averages over the generated lattice configurations, which are related to the path integral. In NNQS, observables are calculated directly as expectation values of operators with respect to the trial wavefunction. 
        
    \item Hyperparameters: NNQS involve a large number of hyperparameters including the number of layers in the network, the number of neurons per layer, the type of activation functions, the initialisation methods and the learning rate. The selection and tuning of these hyperparameters can greatly influence the learning dynamics and the accuracy of the models. The key hyperparameters in the MC approach are the number of samples and the choice of update algorithm. Although the update algorithm affects the rate of convergence, the overall method is quite robust. 
\end{itemize}
\noindent Another point to emphasise are the trade-offs between MC and NNQS in situations where both can be used. One swaps a higher-dimensional sampling problem for a lower-dimensional sampling problem with an additional optimisation problem. Naively, this optimisation problem may eat up any efficiency gain in reducing the dimension of the lattice. However, this optimisation problem can be put on hardware designed for large-scale machine learning tasks, such as clusters of GPUs, while utilising existing software libraries which implement automatic differentiation, etc. In addition, one also benefits from ``transfer learning'' -- the wavefunction for a lattice system found for one value of a coupling will give a good starting point for the wavefunction at a nearby coupling. Thanks to this, one does not have to solve the optimisation problem from scratch when scanning over couplings or probing phase diagrams. The differences between the two methods discussed above are summarised in Table \ref{tab:mc_vs_nnqs}.

\subsection{L-CNNs}\label{sec:LCNN}

The neural network architecture that we will use to approximate the ground state of a $\bZ_{N}$ lattice gauge theory is a so-called ``lattice gauge-equivariant convolutional neural network'', or L-CNN. This particular architecture was introduced by Favoni et al.~in \cite{Favoni:2020reg} as a way to approximate a large class of gauge-equivariant or gauge-invariant functions of a lattice system.\footnote{See also \cite{Luo:2020stn,Luo:2022jzl} for an alternative gauge-equivariant architecture.} In our case, since the network should approximate the ground-state wavefunction of the system, we want the network output to be gauge invariant.

\begin{figure*}
\includegraphics{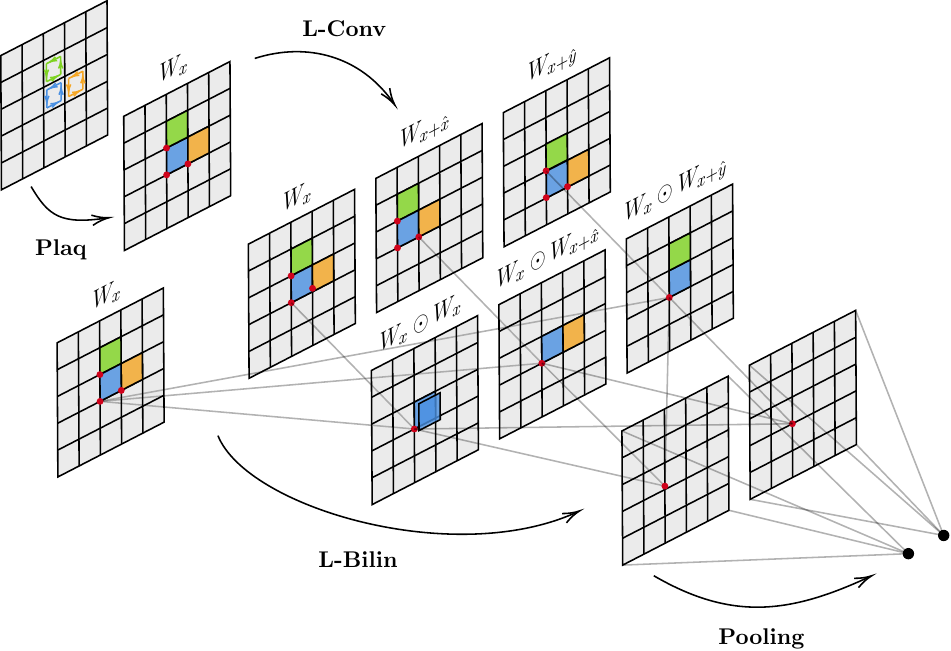}
\caption{A schematic of an L-CNN with a single \textbf{L-CB} layer with $(N_\text{out},K)=(2,1)$ and no \textbf{Dense-SELU} layers. Here, the input is a configuration of link variables $\mathcal{U}=\{U_{x,\mu}\}$. The first layer, \textbf{Plaq}, converts the link variables into plaquette variables and stores them as $\mathcal{W}=\{W_x\}$ in a single channel. The \textbf{L-Conv} layer then parallel transports the plaquette variables by up to one lattice site in the positive $x$ and $y$ directions, giving a new set of plaquette-like variables $\mathcal{W}'=\{W_x, W_{x+\hat{x}}, W_{x+\hat{y}}\}$ with three channels. The sets $\mathcal{W}$ and $\mathcal{W}'$ are then multiplied by element-wise multiplication (denoted by the Hadamard product $\odot$), giving $W_x \odot W_x$ (squares of $1\times 1$ loops), $W_x \odot W_{x+\hat{x}}$ ($2\times1$ loops), and $W_x \odot W_{x+\hat{y}}$ ($1\times2$ loops). Since $N_\text{out}=2$, one then takes linear combinations of these (determined by the weights $\alpha_{ijk}$) to give two output channels. Finally, a global pooling layer averages each of these channels over the lattice, giving two neurons which can be processed further. To help the reader follow variables through the network, we have highlighted three $1\times1$ plaquettes in blue, green and orange. Red dots indicate the lattice points to which the plaquette variables are associated. For example, the action of $W_x \odot W_{x+\hat{x}}$ equivariantly multiplies the blue plaquette by the orange plaquette (since they are associated to the same lattice point), resulting in a $2\times1$ plaquette.}
\label{fig:network}
\end{figure*}

An L-CNN is made up of constituent ``layers''. Depending on their construction, the layers can do a variety of operations, including gauge-equivariant convolutions and multiplications, and acting with activation functions. For our purposes, we will need three of these layers, namely a plaquette layer, a convolution layer and a bilinear layer. At each layer, we keep track of two sets of data. The first is the set of link variables $\mathcal{U}=\{U_{x,\mu}\}$, which transforms non-locally under gauge transformations, as in \eqref{eq:link_gauge_transform}. The second set transforms locally under gauge transformations, as in \eqref{eq:plaqs}. We refer to these collectively as $\mathcal{W}=\{W_{x,i}\}$. Here, $i$ is a ``channel'' index which allows us to associate multiple $\mathcal{W}$ elements to the same lattice site at $x$. The plaquette variables $\{U_{x,\mu\nu}\}$ are an example of these with a single channel; an L-CNN will naturally construct more general quantities which transform in the same way. We give a schematic diagram of an L-CNN in Figure \ref{fig:network}. 

Each layer of the L-CNN can be thought of as acting on the pair $(\mathcal{U},\mathcal{W})$. The initial input to the network is the set of link variables $\mathcal{U}$, describing a gauge field configuration on the lattice, while $\mathcal{W}$ is initially empty. The first layer carries out ``preprocessing'', generating the plaquette variables from the link variables. This is done via a \textbf{Plaq} layer:
\begin{equation}
\text{\textbf{Plaq}}\colon U_{x,\mu}\mapsto U_{x,\mu\nu}.
\end{equation}
The plaquette variables are then stored in $\mathcal{W}$. As in \cite{Favoni:2020reg}, to reduce redundancy, we generate only the anticlockwise plaquettes (those with positive orientation in higher dimensions).

The next layer allows the parallel transport of objects stored in $\mathcal{W}$ from one lattice site to a neighbouring site. This is implemented as a convolutional layer, \textbf{L-Conv}, given explicitly by
\begin{equation}
\text{\textbf{L-Conv}}\colon (U_{x,\mu},W_{x,i})\mapsto\sum_{j,\mu,k}\omega_{ij\mu k}U_{x,k\mu}W_{x+k\mu,j}U_{x,k\mu}^{\dagger},
\end{equation}
where $\omega_{ij\mu}\in\bC$ are the ``weights'' or parameters of the convolutional layer, $i$ and $j$ run over the number of output and input channels respectively, and $\mu\in\{\hat{x},\hat{y}\}$ for a $d=2$ lattice. The index $k$ runs over $0,\dots,K$, where $K$ is an integer which determines the maximum lattice distance to translate the $\mathcal{W}$ quantities, or equivalently the kernel size of the convolution.\footnote{We restrict to non-negative shifts along the lattice.} 

Finally, we need a layer which multiplies two sets $\mathcal{W}$ and $\mathcal{W}'$ in an equivariant manner. This is done via an \textbf{L-Bilin} layer:
\begin{equation}\label{eq:L-Bilin}
\text{\textbf{L-Bilin}}\colon(W_{x,i},W_{x,i'}')\mapsto\sum_{j,k}\alpha_{ijk}W_{x,j}W'_{x,k},
\end{equation}
where $\alpha_{ijk}\in\bC$ are weights, $j$ and $k$ run over the number of input channels for $\mathcal{W}$ and $\mathcal{W}'$ respectively, and $i$ runs over the number of output channels. Since this layer multiplies locally transforming variables at the same position $x$, the output is again locally transforming, so that the layer output is gauge equivariant. For $\bZ_{N}$ gauge theories, the gauge group is abelian and so the link variables are simply phases. Thanks to this, the traced and untraced Wilson loops are equivalent, and any function of the $\mathcal{W}$ variables is automatically gauge invariant.

In practice, prior to multiplying, $\mathcal{W}$ and $\mathcal{W}'$ are extended by including the Hermitian conjugate of all their elements, and the unit matrix $\id$ at each lattice site. A little thought should convince the reader that this allows the layer to include a bias and act as a residual module (i.e.~the output also contains a linear combination of the inputs)~\cite{ResNet}. As in \cite{Favoni:2020reg}, we combine \textbf{L-Conv} and \textbf{L-Bilin} into a single \textbf{L-CB} layer with a single set of trainable weights. A choice of \textbf{L-CB} layer is then fixed by a choice of $(N_{\text{out}},K)$, i.e.~the number of output channels and the kernel size of the convolution.

The power of an L-CNN is in the fact that by stacking \textbf{L-CB} layers, one can construct arbitrary (untraced) Wilson loops, and so approximate any gauge equivariant function. For example, after a \textbf{Plaq} layer, one has all $1\times1$ Wilson loops. Following this with an \textbf{L-CB} layer, the output includes linear combinations of $1\times2$ and $2\times1$ Wilson loops (and $1\times1$ loops and squares of $1\times1$ loops when the $\mathcal{W}$ variables are extended by the unit matrix). If the number of output channels is large enough, in principle, one can capture all possible Wilson loops of area two and below. With another \textbf{L-CB} layer, the output includes loops up to area four. This also makes clear that L-CNNs exploit locality on the spatial lattice, with nearby Wilson loops more likely to be multiplied together compared with distant loops. 

Clearly, the number of possible Wilson loops grows very quickly with area, so quickly that one cannot hope to optimise a variational ansatz constructed by simply taking combinations of all loops. Instead, by restricting the size of the output channels, an L-CNN works with a much smaller number of combinations of loop variables. During training, the network then determines which combinations to keep within this much smaller subspace. It is this restriction that ensures an L-CNN variational ansatz has a sub-exponential number of parameters. For example, for $\bZ_2$ gauge theory on a $10\times10$ lattice, there are $2^{2\times{10}^2}\approx {10}^{60}$ possible lattice configurations, and so modeling the wavefunction as a ``look-up table'' that assigns an amplitude to each of these is clearly intractable.\footnote{This is obviously an overcounting, as the ground-state wavefunction depends only on gauge-invariant data and should also be invariant under translations, etc., but the rough level of complexity is what the reader should take away.} Instead, using an L-CNN with seven \textbf{L-CB} layers with $(N_{\text{out}},K) = (4,2)$ leads to a variational ansatz with approximately ${10}^4$ parameters. As we will see, this network is sufficient to accurately capture the physics of the ground state, and so an L-CNN clearly gives an efficient encoding of the wavefunction.

\subsection{NetKet implementation}\label{sec:netket}

We have implemented this network architecture using NetKet~\cite{Vicentini:2021pcv}, a machine-learning framework for many-body quantum physics. NetKet is built on top of JAX~\cite{JAX}, a framework for Python which allows automatic differentiation and GPU acceleration, with its neural network components implemented using Flax~\cite{FLAX}. The network output is taken to be $\log\Psi_\theta$. Thanks to the complex weights of the network, the output is complex, so can accommodate a wavefunction with non-trivial phase structure. Schematically, as a functional, the wavefunction is given by the composition of the following layers
\begin{equation}
\log\Psi_\theta\equiv\text{\textbf{Dense}}\circ\dots\circ\text{\textbf{Dense-SELU}}\circ\text{\textbf{Pooling}}\circ\dots\circ\text{\textbf{L-CB}}\circ\text{\textbf{Plaq}},
\end{equation}
where the final layers are fully connected dense layers, each followed by a scaled exponential linear unit layer (\textbf{Dense-SELU}), and a dense layer with a \emph{single} neuron (\textbf{Dense}). Each \textbf{L-CB} layer is labeled by a choice of $(N_\text{out},K)$, the number of output channels and the kernel size, while each \textbf{Dense-SELU} layer is fixed by $N_\text{feat}$, a choice of the number of neurons or ``features'' in the dense layer.

The network is trained using the in-built features of NetKet. Specifically, training attempts to minimise $E_\theta$, the expectation value of the Hamiltonian in the variational state. It does this using automatic differentiation to compute the derivative (gradient) of $E_\theta$ with respect to the parameters (weights) of the network, and then stochastic gradient descent to move in the steepest direction of lower energy. In addition, NetKet includes the option to precondition the gradient -- we use the quantum geometric tensor in all numerical experiments in this paper, with the resulting dynamics known as stochastic reconfiguration.\footnote{See, for example, \cite[Section 4.1]{Vicentini:2021pcv} for a discussion of this.} We further elucidate the training process below.

\subsubsection{Sampling}

Gradient descent requires the calculation of $E_\theta$ and its gradient with respect to the parameters of the network at each training step. In principle, computing these quantities requires summing over the Hilbert space of the system. Since the Hilbert space of gauge field configurations is too large to sum over exactly, one must instead use stochastic gradient descent with an estimate of the gradient. The expectation values of observables and their gradients with respect to the network parameters are computed using a representative sample of configurations which approximate the full sum over the Hilbert space. Here, representative means that a configuration $\mathcal{U}$ is sampled according to its probability $|\Psi_\theta(\mathcal{U})|^2$. These are selected via a standard local Metropolis algorithm. As $\Psi_\theta$ converges to the ground-state wavefunction, the sampling becomes better at reproducing the sum over the Hilbert space.

Obviously, in order to actually move in the direction of decreasing energy, one needs reasonably accurate estimates of the gradient of the energy. Since this gradient is calculated approximately by summing over a sample of lattice configurations, one might worry that this gradient (or the energy itself) cannot be estimated to sufficient accuracy without using a very large number of samples. However, the Hamiltonian is a particularly well-behaved observable as it satisfies the ``zero-variance property''.\footnote{See, for instance, the discussion in \cite{10.1063/1.1621615}} Given a trial wavefunction $\Psi_\theta$, one can consider both the variational error in the energy, $E_\theta - E_0$, and the variance, $\langle (H - E_\theta)^2\rangle_\theta$. One can show that both the variational error and the variance are second order in the difference between the trial wavefunction $\Psi_\theta$ and the true ground-state wavefunction $\Psi_0$. In the limit where $\Psi_\theta = \Psi_0$, both the error and the variance vanish, so that $E_\theta = E_0$. Moreover, since the variance determines the statistical error, the error in computing the gradient of the variational energy due to Monte Carlo sampling also decreases as $\Psi_\theta$ approaches $\Psi_0$. Thanks to this, one does not need a large number of samples when estimating the energy or its gradient. 

For the experiments described in this paper, each stochastic gradient descent step is evaluated using 4096 configurations. The number of sweeps (the number of Metropolis steps taken before returning a sample, i.e.~the subsampling factor of the Markov chain) is chosen to be equal to the number of degrees of freedom of the Hilbert space. This is simply the number of links, so that for an $L\times L$ lattice, $2L^2$ sweeps are made. Combined with a warm-up phase, we found this to be sufficient to ensure reasonably small correlation of the Monte Carlo chains and an acceptable autocorrelation time. When approaching critical couplings, an increase in the number of sweeps was needed.

\subsubsection{Initialisation}

Since \textbf{L-CB} layers are multiplicative, when using deep networks, it is essential to properly initialise the network weights. A little thought should convince the reader that for a deep network, if the weights are initially too small or too large, one will quickly run into vanishing or exploding gradients. For fully connected or convolutional networks with standard activation functions, there are analytic results for choosing a good initialisation. Without similar results for multiplicative networks, such as an L-CNN, the best one can do is to choose the initialisation empirically. Following \cite{2015arXiv151106422M}, one can do this via ``layer-sequential unit-variance'' initialisation. The idea is that one starts with some distribution of weights for each layer, with known standard deviations. One then proceeds, layer by layer, changing the standard deviation so that the output of each layer has the same variance as the previous layer. In this way, the gradient should not vanish nor explode. We have implemented this for all the networks that we discuss. We find that this is essential for ensuring that deep networks do not immediately diverge, nor take a long time to begin training.

\subsubsection{Training details}

For all the networks in this paper, we used stochastic gradient descent together with a preconditioning of the gradient via the quantum geometric tensor. In the variational Monte Carlo community, this is known as stochastic reconfiguration~\cite{cond-mat/9803107,cond-mat/0009149}. The update rule for the weights $\theta$ of a variational state $|\Psi_\theta\rangle$ is
\begin{equation}
    \theta \to \theta - \eta\, G^{-1} \nabla_\theta E_\theta,
\end{equation}
where $\eta$ is the learning rate and $G^{-1}$ is the (pseudo-)inverse of the quantum geometric tensor. This tensor (also known as the quantum Fisher matrix) is the metric tensor induced by the Fubini--Study distance between pure quantum states. The resulting dynamics is akin to what is known as ``natural'' gradient descent in the machine-learning community~\cite{10.1162/089976698300017746}, where $G^{-1}$ takes into account that the space of states has non-trivial geometry~\cite{1910.11163}. In practice, $G$ is often ill-conditioned, and so a small diagonal shift proportional to the identity matrix is often added before inversion. For large diagonal shifts, the identity matrix will dominate $G$, leading to standard stochastic gradient descent with the Euclidean metric on weight space. This should still converge to the ground state, though may be much slower than choosing an optimally small value of the shift.

In our experiments, the learning rate was initially set to $0.03$ and reduced via cosine decay over 500 iterations. Stochastic reconfiguration was implemented following the ``MinSR'' algorithm~\cite{2302.01941, 2310.05715} using NetKet's experimental \texttt{VMC\_SRt} driver with a diagonal shift of ${10}^{-4}$. Double (FP64) precision was used for all experiments, as we found this led to more stable training of the multiplicative layers that make up an L-CNN.

As discussed in \cite{1503.07045}, the variance of the energy is useful for tracking the convergence of a variational method.\footnote{See also \cite{2302.04919} for the ``V-score'', which can be thought of as a system-size agnostic definition of the accuracy of a variational state.} Given an approximate eigenstate with energy $E=\langle H\rangle$ and $\Delta E=\sqrt{\op{var}H}=\sqrt{\langle H^{2}\rangle-\langle H\rangle^{2}}$, there will be an exact energy eigenvalue $E_{\text{exact}}$ within $\Delta E$ of the energy $E$, $|E-E_{\text{exact}}|\leq \Delta E$~\cite{10.1073/pnas.20.9.529,PhysRev.53.199.2}. The variance thus gives an upper bound on how far an approximate state is from an exact energy eigenstate, and thus can be used as a stopping condition.\footnote{One might worry that training might become stuck at an excited state, and that the variance cannot be used to discriminate between this and the true ground state. Fortunately, stochastic reconfiguration is excellent at driving the network to the ground state, after which the variance stopping condition can be trusted.} Training is stopped once the network has converged, which is indicated by $E_\theta$ no longer changing and the variance of $H$ becoming sufficiently small.\footnote{In practice, we use the sample variance of $E_\theta$ to estimate the variance of $H$ while training.} Upon convergence, the resulting trained network should then approximate the ground-state wavefunction.

\subsubsection{Transfer learning}

One of the advantages of computing the wavefunction of a quantum system, rather than using path-integral methods to compute observables directly, is the possibility of employing transfer learning. This relies on the fact that on a finite lattice, small changes in the parameters of the Hamiltonian should lead to only small changes in the ground-state wavefunction. 
We exploit this when performing scans over the coupling: by starting with the network weights corresponding to a nearby, previously learned wavefunction, the new NNQS is already relatively close to the sought-for ground state. This also helps with stability near to critical values of the of coupling, since the NNQS starts from a nearby wavefunction (in state space), rather than trying to converge from a generic state. We find this form of transfer learning helps the network to converge and greatly decreases overall training time. 

There is also a second kind of transfer learning that we could take advantage of (though we did not in our experiments). Since an L-CNN is a convolutional network, it implements weight sharing for different lattice sites. Said differently, the trainable weights in an \textbf{L-CB} layer are encoded in a rank-three tensor with entries $\alpha_{ijk}\in\bC$, where the indices run over the output and input channels. The number of these channels does not depends on the hyperparameters $(N_\text{out},K)$ for each layer, but does not depend on the size of the underlying lattice. This means that one can train an L-CNN on a small lattice, and then transfer the network weights to an L-CNN for a larger lattice. Presumably, this would provide a good starting point for learning the wavefunction on the larger lattice.

\subsubsection{Calculating observables}

At the end of training, one has a NNQS which approximates the ground-state wavefunction of the system. With this in hand, one would like to compute other observables in order to probe various aspects of its physics. Unlike the Hamiltonian, general observables do not enjoy the zero-variance property. This has important consequences for computing accurate expectation values. Given an observable $\mathcal{O}$, the variational error $\langle \mathcal{O}\rangle_\theta - \langle\mathcal{O}\rangle_0$ is no longer quadratic in the difference $\Psi_\theta - \Psi_0$, but changes linearly. Moreover, the variance $\op{var}_\theta \mathcal{O} = \langle (\mathcal{O} - \langle \mathcal{O}\rangle_\theta)^2\rangle_\theta$ remains order one (effectively because the exact ground-state does not have to be an eigenstate of $\mathcal{O}$). Thus, the statistical fluctuations of $\langle\mathcal{O}\rangle_\theta$ can be large, requiring many samples to reduce the standard error in the estimate, which naively goes as the square root of $\op{var}_\theta\mathcal{O} / n_\text{samples}$ modulo the effects of autocorrelation. 

\subsection{Hilbert space sector of the NNQS}

The `t Hooft loop operators along the two one-cycles of the torus are the generators of the $\mathbb{Z}^{(1)}_{N}$ one-form global symmetry and commute with the Hamiltonian. Therefore, the Hilbert space can be decomposed into selection sectors based on the eigenvalues of the `t Hooft loop operators~\cite{Kitaev:2005hzj}. Within each sector, the states can be transformed from one to another by acting with contractible Wilson loop operators comprised of gauge field on links. This is in contrast to states belonging to different sectors that cannot be transformed into each other in this manner. Given our discussion of ground-state degeneracies in the deconfined phase in Section \ref{sec:crit_behaviour}, one might wonder which sector the NNQS belongs to. Is the neural network finding some superposition of the approximately degenerate ground states? In fact, as we now show, the network architecture ensures that one is always in the trivial `t Hooft loop sector. Therefore, the NNQS recovers the true ground states in both the confined and deconfined phases. To see this, as we review in Appendix \ref{app:WL}, recall that the expectation value of an observable $\mathcal{O}$ can be expressed as
\begin{equation}
\langle\mathcal{O}\rangle=\frac{\langle\Psi|\mathcal{O}|\Psi\rangle}{\langle\Psi|\Psi\rangle}=\sum_{\mathcal{U}}\frac{|\Psi(\mathcal{U})|^{2}}{\langle\Psi|\Psi\rangle}\left(\sum_{\mathcal{U}'}\frac{\Psi(\mathcal{U}')}{\Psi(\mathcal{U})} \langle\mathcal{U}|\mathcal{O}|\mathcal{U}'\rangle\right).
\end{equation}
The summation over all possible field configurations $\mathcal{U}$ is usually approximated by sampling using Markov chain Monte Carlo (MCMC). The sum over $\mathcal{U}'$ inside the parentheses receives contributions only from ``connected configurations'', that is, configurations where $|\mathcal{U}\rangle$ and $\mathcal{O}|\mathcal{U}'\rangle$ have non-zero overlap. In the case where $\mathcal{O}=\Theta_x$ is a local generator of gauge transformations on the lattice, there is only one connected configuration, and $\Psi(\mathcal{U})$ and $\Psi(\mathcal{U}')$ must be equal since the network used to calculate $\Psi$ is gauge invariant by construction. The expectation value then reduces to an average of $1$'s, giving $\langle\mathcal{O}\rangle=1$, implying that the wavefunction is gauge invariant.

A similar argument applies when $\mathcal{O}$ is taken to be any of the `t Hooft loop operators that wrap the torus. Focusing on $N=2$ for concreteness, one has $\mathcal{O}\in\{V_{x},V_{y}\}$. These operators commute with all contractable Wilson loops, and so they do not change the expectation values of Wilson loops at all:
\begin{equation}
    \langle \mathcal{U}'|W|\mathcal{U}'\rangle = \langle \mathcal{U}|\mathcal{O}^\dagger W\mathcal{O}|\mathcal{U}\rangle=\langle \mathcal{U}|\mathcal{O}^\dagger \mathcal{O}W|\mathcal{U}\rangle=\langle \mathcal{U}|W|\mathcal{U}\rangle.
\end{equation}
Since our variational wavefunction $\Psi$ is built from contractable Wilson loops of all sizes, it is invariant under $V_x$ and $V_y$ by construction. Therefore, we always have $(V_x,V_y)=(1,1)$, which are the quantum numbers of the true ground state (and not of the approximately degenerate states). In order to explore other sectors, one can simply conjugate the Hamiltonian with appropriate combinations of the $V_{x}$ and $V_{y}$ operators.

\begin{figure}
\includegraphics{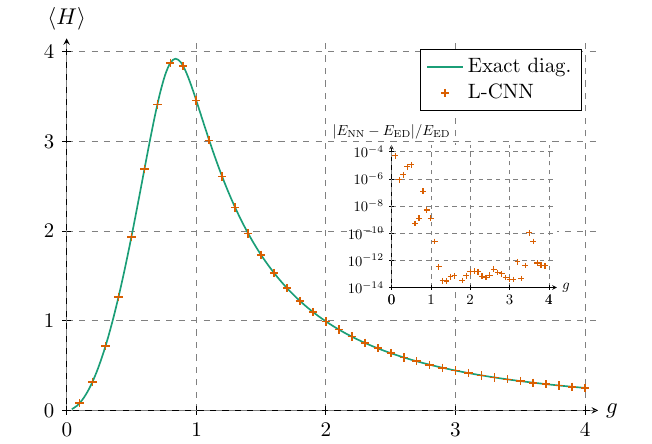}
\caption{Expectation value of $\langle H\rangle$ for $\protect\bZ_{2}$ gauge theory \eqref{eq:hamiltonian_Zn} on a $2\times2$ lattice as a function of coupling $g$. The ground state was computed using an L-CNN with six \textbf{L-CB} layers with hyperparameters $(N_\text{out},K)=(3,1)$. The networks were trained for up to 100 iterations with $4096$ samples, though they converged within 25 or so iterations. We trained the networks from larger to smaller values of the coupling $g$, using the previously trained network as the starting point. The solid teal line is the exact diagonalization result, computed using NetKet. The inset shows the fractional difference between the ground state energies calculated using the neural network and exact diagonalization.}
\label{fig:energy_Z2_2d}
\end{figure}

\section{\texorpdfstring{$\mathbb{Z}_2$}{Z2} gauge theory in \texorpdfstring{$2+1$}{2+1} dimensions}\label{sec:Z2}

We now turn to the study of $\mathbb{Z}_2$ lattice gauge theory in $2+1$ dimensions~\cite{Wegner:1971app,Fradkin:1978th,Elitzur:1979uv,Horn:1979fy,Trebst:2006ci,Tupitsyn:2008ah}. There are surprisingly few direct numerical studies of the ground state of $\bZ_2$ lattice gauge theories, mainly due to the requirement of gauge invariance. Instead, the literature has focused either on path-integral Monte Carlo or simulating the dual spin system. Our focus will be on finding the ground-state wavefunction itself, allowing us to compute the ground-state energy as a function of coupling, to identify the critical coupling and the confined-deconfined phases, and to calculate estimates for the critical exponents that characterise the conformal field theory governing the phase transition. We also investigate the potential between two static charges on the lattice.

As reviewed in \cite{Rayhaun:2021ocs}, $\bZ_2$ gauge theory at zero temperature is known to have two phases: an ordered (deconfined) phase and a disordered (confined) phase. These phases are distinguished by the behaviour of the order and disorder parameters, given by expectation values of Wilson loop and `t~Hooft string operators. $\bZ_2$ gauge theory is dual to a classical Ising model in $2+1$ dimensions, and therefore the second-order phase transition at $\gc$ is in the universality class of the three-dimensional gauged Ising CFT. 

The ordered phase is expected to appear for couplings below a critical coupling $\gc$, and is characterised by perimeter-law decay for Wilson loops. The $\bZ_2^{(1)}$ one-form symmetry is broken in this phase. The slow decay of Wilson loops indicates that the ground state of the theory is dominated by these operators. One then says that the corresponding electric flux lines are ``condensed'' and the theory is deconfined. Above the critical coupling, the theory is in a disordered phase, with the Wilson loop expectations following an area-law decay. The `t~Hooft string expectations are constant (independent of distance) due to the condensation of magnetic monopoles. The $\bZ_2^{(1)}$ one-form symmetry is unbroken and the theory is confined.

\subsection{Ground-state energies}

As a first test of the L-CNN, we compute the ground-state energy of pure $\bZ_{2}$ lattice gauge theory in $2+1$ dimensions. In Figure \ref{fig:energy_Z2_2d}, we plot the expectation value of the lattice Hamiltonian \eqref{eq:hamiltonian_Zn} in the ground state as a function of coupling $g$ for a $2\times2$ lattice. We see excellent agreement with the energy calculated by exact diagonalization.

\begin{figure}
\includegraphics{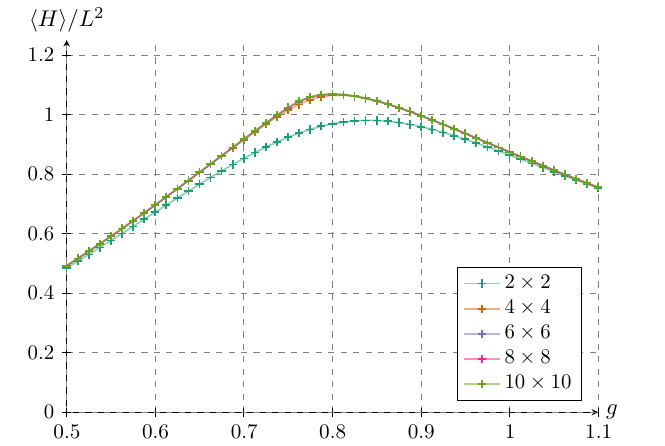}
\caption{Expectation value $\langle H\rangle$ for $\protect\bZ_{2}$ gauge theory \eqref{eq:hamiltonian_Zn} for varying lattice sizes as a function of coupling $g$, focused on the region around the critical coupling. The networks were trained for up to 2000 iterations with $2048$ samples. Training was stopped early when the variance of the energy per site stabilised to a value of $1.25\times{10}^{-4}$ or less. The expectation values were then calculated using $32\times4096$ Monte Carlo samples. We trained the networks from larger to smaller values of the coupling $g$, using the previously trained network as the starting point.}
\label{fig:Z2_LxL_energy}
\end{figure}

\begin{figure}
\includegraphics{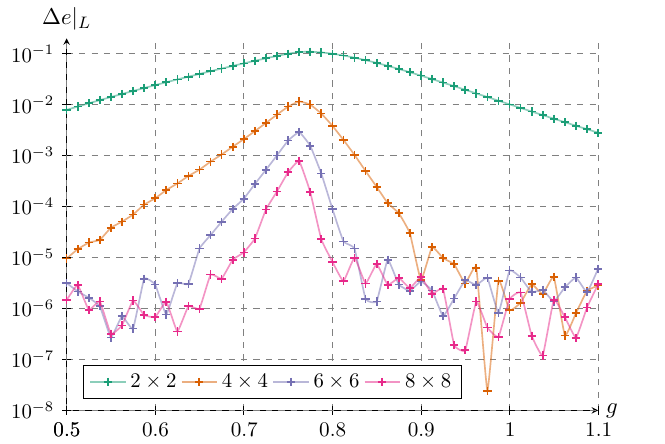}
\caption{The difference between the calculated ground-state energy per lattice site for $L=10$ and varying $L$, $\Delta e|_L = \left|E_{L=10}/{10}^2 - E_L/L^2\right|$, as a function of coupling $g$. Taking the $L=10$ result to be our ``continuum'' values, we see that the energies quickly approach this, even for smaller lattice sizes.}
\label{fig:Z2_LxL_energy_diffs}
\end{figure}

In Figure \ref{fig:Z2_LxL_energy}, we plot the ground-state energy per lattice site, $\langle H\rangle / L^2$, for couplings in the range $g\in[0.5,1.1]$ with lattice sizes $L=2,\dots,10$. By eye, one sees that the $L=10$ values are likely already very close to the continuum energy per site and that even relatively small lattices provide a good estimate, with only $L=2$ showing large deviations. We quantify this further in Figure \ref{fig:Z2_LxL_energy_diffs}, which shows the difference between the ground-state energy per lattice site for $L=10$ and smaller values of $L$. Other than for $L=2$, away from $g\in[0.7,0.8]$, the estimates agree to better than ${10}^{-4}$ to ${10}^{-6}$. It is also interesting to observe that the differences are maximised for $g\approx 0.76$ which, as we will see, is in the vicinity of the phase transition.

\subsection{Searching for the critical point and phase structure}
The critical coupling $\gc$ for the $\bZ_2$ gauge theory has not been directly computed previously. From \cite[Figure 5]{Luo:2020stn}, the behaviour of the string tension on a $12\times12$ lattice suggests $\gc\approx 0.74$.\footnote{The critical coupling is given in \cite{Luo:2020stn} as $h \approx 0.3$. Examining their Hamiltonian, one finds that the $h$ coupling is related to ours via  $h_\text{there}^{1/4} = g_\text{here}$.} However, this comes from eyeballing the area-law scaling of Wilson loops and is accurate to, at best, one significant figure. Later work suggested a transition around $\gc\approx0.76$ from considerations of `t Hooft string expectations and derivatives of the energy~\cite{Luo:2021ccm}. Instead, the most accurate identifications of the critical coupling come from a Monte Carlo analysis of the dual spin system, the quantum transverse-field Ising model. The authors of \cite{PhysRevE.66.066110} simulate an anisotropic limit of the Ising model on a $2+1$-dimensional lattice, equivalent to the two-dimensional quantum transverse-field Ising model on a square lattice. Their result is that $\gc = 0.757051$.\footnote{Our coupling is related to theirs via $t_\text{there}^{-1/4} = g_\text{here}$.}

\begin{figure}
\includegraphics{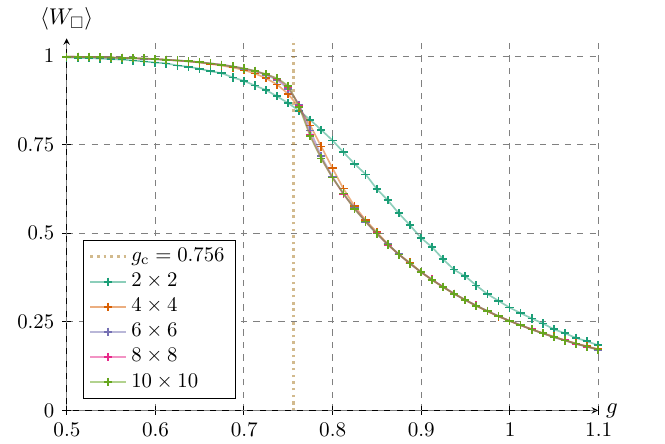}
\caption{The lattice average of the single-plaquette Wilson loop operator $W_\square$ in $\bZ_2$ gauge theory as a function of coupling $g$ near the critical point for varying lattice sizes. This corresponds to the expectation value of the magnetic flux energy per site. The dotted yellow line indicates the position of the critical coupling at $\gc=0.756$, as determined by our analysis in Section \ref{sec:BST_coupling}.}\label{fig:Z2_wilson_loop}
\end{figure}

\begin{figure}
\includegraphics{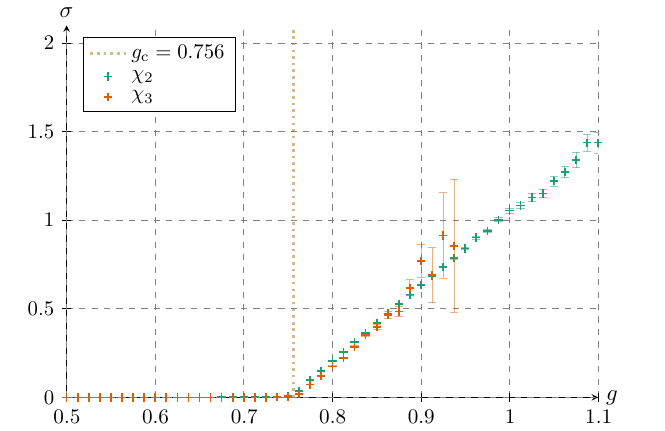}
\caption{String tension estimated from the Creutz ratio $\chi_l$ for $l=2,3$ on a $10\times10$ lattice using $2^{20}\approx 1\text{M}$ Monte Carlo samples. The error bars shown only take into account Monte Carlo errors. For $g>g_{\text{c}}$, Wilson loops decay rapidly with size. Estimates of $\sigma$ are then challenging due to ratios of very small numbers, as can be seen from the noisier signal for $g>1$ with $l=2$ and $g>0.85$ with $l=3$. In particular, above $g=0.9375$, the estimates of $\chi_3$ are unreliable, with the standard error becoming larger than the observable. As such, we have not shown these points.}\label{fig:Z2_creutz}
\end{figure}

A first attempt at locating the phase transition might proceed by looking for signs of the confinement-deconfinement transition using the order parameter of the theory. Recall that though the order parameter is the Wilson loop operator, confinement is not cleanly characterised by its expectation value, but by a change from perimeter- to area-law decay. Indeed, as can be seen in Figure \ref{fig:Z2_wilson_loop}, $\langle W_\Box\rangle$ is non-zero for all finite couplings on a finite-size lattice. Instead, it is the string tension that displays the usual behaviour of an order parameter -- zero in the  in the area-law decay of the loops. As discussed in Section \ref{sec:crit_behaviour}, given the ground-state wavefunction, the string tension can be estimated using the Creutz ratio. In Figure \ref{fig:Z2_creutz}, we show the string tension $\sigma$ estimated using $\chi_2$ on a $10\times10$ lattice. In the small-coupling regime, sufficiently far from the phase transition, the string tension is zero to within Monte Carlo errors. As we approach $g\approx 0.76$, the expected critical coupling, we see that the string tension becomes non-zero and has approximately linear growth with coupling. Using a rough linear fit to this regime, one finds $\sigma(g) = -3.48 + 4.59 g$. Extrapolating back to zero string tension implies $\gc\approx 0.758$, which is already in good agreement with Monte Carlo results from the dual spin system. For larger couplings, $g>0.95$, Monte Carlo errors become of the same order as the quantities appearing in the Creutz ratio, leading to a noisier signal. Figure \ref{fig:Z2_creutz} also shows the string tension estimated using $\chi_3$. Since larger Wilson loops should be less affected by finite-size corrections, this should give a better estimate of $\sigma$, though we also observe that the signal quickly becomes dominated by Monte Carlo errors in the confined phase.

\begin{figure}
\includegraphics{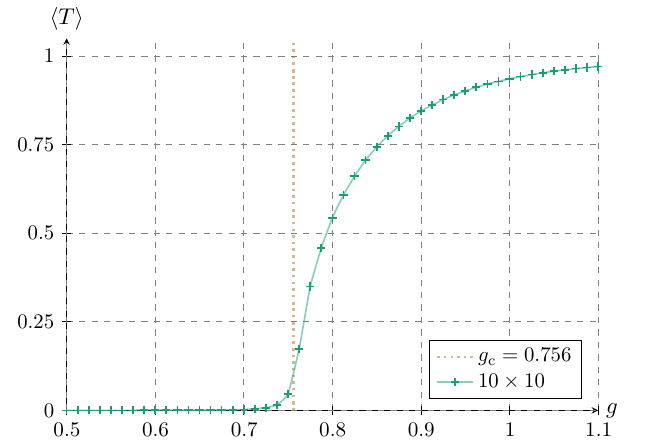}
\caption{The lattice average of the `t~Hooft string operator in $\bZ_2$ gauge theory as a function of coupling $g$ near the critical point for a $10\times10$ lattice. The string operator is of length five in the $x$ direction.}\label{fig:Z2_tHooft_10x10}
\end{figure}

Next, we can look for evidence of the phase transition in the disorder parameter, i.e.~the `t~Hooft string. In the ordered/deconfined phase, expectation values of these string operators decay with the distance between the ends of the string, while in the disordered/confined phase, they should be constant. In Figure \ref{fig:Z2_tHooft_10x10}, we plot the disorder parameter as a function of coupling for monopoles a distance five apart on a $10\times10$ lattice. We see that below the critical coupling, the expectation values decays quickly to zero, while near to the transition it grows rapidly, eventually approaching one. These results should be compared with those in \cite[Figure 11 (b)]{Luo:2021ccm}, which display a similar, though much noisier, transition.

At this point, we have demonstrated the presence of confined and deconfined phases, with the transition occurring at around $0.76$. We would now like to identify the critical couplings more accurately and to establish whether the transition is first-order or continuous. We will see that it is continuous, in agreement with the literature, and then go on to calculate the critical exponents that characterise the conformal field theory that governs the critical point.

\subsection{Critical exponents and finite-size scaling}\label{sec:exponents}
\begin{figure*}
\hfill{}
\includegraphics[trim={32.10236pt 0 32.10236pt 0}]{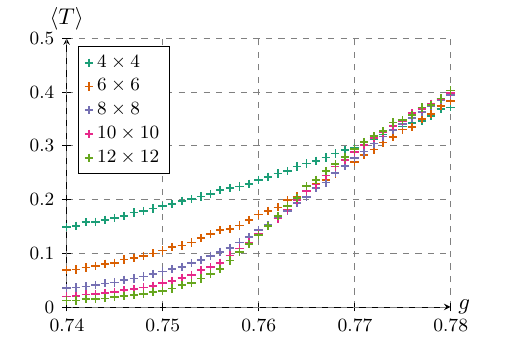}%
\hfill{}
\includegraphics[trim={27.75078pt 0 25.575pt 0}]{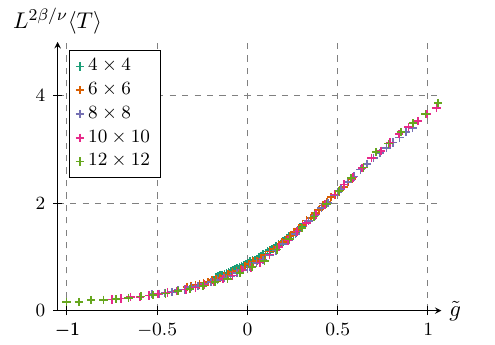}%
\hfill{}
\caption{Lattice average of the `t~Hooft string for varying lattice size and distance $L/2$ between ends of `t~Hooft string. The left figure shows the unscaled expectation values as a function of coupling, spaced equally over the range $g\in[0.74,0.78]$ around the critical coupling. The right figure shows curve collapse for the rescaled observable $L^{2\beta/\nu} T$, with the rescaled coupling $\tilde{g}=L^{1/\nu}(g-\gc)/\gc$. The critical coupling and exponents are given by $\gc=0.7546$, $\beta=0.326$ and $\nu=0.630$. Note that, since $\tilde{g}$ depends on $L$, a fixed range of $g$ values is mapped to a varying range of $\tilde{g}$ values, as can be seen in the figure.}\label{fig:tHooft_collapse}
\end{figure*}

The discontinuities and singular behaviour that characterise phase transitions appear only in the thermodynamic limit, wherein the volume of the system (the lattice size) approaches infinity \cite{Yang:1952be, Lee:1952ig}. This is a result of the fact that for finite system size,  all calculations involving the ground state and path integrals are finite and lead to smooth functions. Away from the critical coupling for the phase transition, there is an exponential decay of two-point correlation functions as $\ee^{-r/\xi}$, where $r$ is the distance between the operator insertions and $\xi$ is the correlation length. For continuous phase transitions, the correlation length diverges as the system approaches the critical point $ \xi = \xi_{0}|g-g_{\text{c}}|^{-\nu}$ with a positive critical exponent $\nu>0$. For a finite-size system of size $L$, the theory of finite-size scaling describes how observables behave in the vicinity of the critical point as a function of the length scale $L$ and the distance from the critical point~\cite{10.1103/PhysRevLett.28.1516}. The key insight of this theory is that, close to the critical coupling $\gc$, the correlation length $\xi$ is comparable to the system size and therefore the microscopic length given by the lattice spacing which governs the range of interaction no longer affects the correlation functions on scales larger than the lattice spacing.

The Ising model has a $\mathbb{Z}_{2}$ global symmetry, which corresponds to flipping all the spins on the lattice. Recall that in the Ising model, one has to tune the temperature to the critical value $T=T_\text{c}$ and the external field to zero to obtain the Ising CFT, which inherits this $\mathbb{Z}_{2}$ global symmetry. Consequently, all operators are either $\mathbb{Z}_{2}$ odd (e.g.~the spin field $\sigma$) or $\mathbb{Z}_{2}$ even (e.g.~the energy density field $\epsilon$) and there are exactly two relevant scalar operators in the CFT~\cite{Rychkov2020}. The $\mathbb{Z}_{2}$-even scalar with the lowest scaling dimension is the energy density field $\epsilon$, and its scaling dimension $\Delta_{\epsilon}$ is related to the critical exponent $\nu$, which characterises the divergence of the correlation length as $T \to T_\text{c}$. The $\mathbb{Z}_{2}$-odd scalar with the lowest scaling dimension is the spin field $\sigma$, which transforms as $\sigma \to -\sigma$ under the $\mathbb{Z}_{2}$ global symmetry and serves as the order parameter for the $\mathbb{Z}_{2}$ symmetry breaking phase transition. The scaling dimension $\Delta_{\sigma}$ is related to the magnetization critical exponent $\beta$, which characterises the vanishing of magnetization as temperature is increased to the critical value. Renormalization group (RG) analysis reveals the precise relation between scaling dimensions of these operators and the critical exponents in $d$-dimensions~\cite{Cardy1996}:
\begin{equation}
    \nu = \frac{1}{d-\Delta_{\epsilon}},\qquad \beta = \frac{\Delta_{\sigma}}{d-\Delta_{\epsilon}}. 
\end{equation}
As $\mathbb{Z}_{2}$ lattice gauge theory can be obtained by gauging the $\mathbb{Z}_{2}$ global symmetry of the Ising model, the theory at the critical coupling $\gc$ is described by the $\mathbb{Z}_{2}$ gauged Ising CFT (Ising$^{\ast}$ CFT). Since the two theories are related by a discrete gauging, the corresponding scaling dimensions and critical exponents are identical. The $\mathbb{Z}_{2}$ odd local operators get projected out during the gauging procedure. However, the $\mathbb{Z}_{2}$ odd $\sigma$ field that is attached to the `t~Hooft string operator remains in the spectrum. Consequently, the order parameter that we employ to diagnose the confinement phase transition and compute the critical exponents $\beta$ and $\nu$ is the lattice average of the expectation value of a `t~Hooft string operator.  

The `t~Hooft string is taken to lie in the $x$-direction with length $L/2$, corresponding to one-half of the lattice size. This choice numerically leads to the smallest variance and, since the length of the string operator scales with the system, it ensures that one obtains a line operator in the $L\to\infty$ limit. Since there are two $\sigma$ fields (monopole operators) at the end of the string operator, the CFT then predicts that in the vicinity of the critical point the expectation value of the operator should vary with $L$ as 
\begin{equation}
    \langle T \rangle \sim L^{-2\Delta_{\sigma}},
\end{equation}
where $\Delta_{\sigma}$ is the scaling dimensions of the $\sigma$ operator attached to the `t~Hooft string in the Ising$^{\ast}$ CFT \cite{Chang:2018iay}.

Combining this with the finite-size scaling hypothesis, the functional dependence can be expressed solely in terms of the dimensionless ratio $\xi/L$ and system size $L$ as
\begin{equation}
    \langle T \rangle = L^{-2\Delta_{\sigma}} f(\xi/L),
\end{equation}
where $f$ is an $L$-independent scaling function. Using the relations $\xi(g) \sim |g-g_{\text{c}}|^{-\nu}$ and $\beta/\nu = \Delta_{\sigma}$, this can be expressed in terms of $g$ using another scaling function $F$:
\begin{equation}\label{eq:scaling_ansatz}
    \langle T \rangle = L^{-{2\beta/\nu}} F\bigl(L^{1/\nu}(g-g_{\text{c}})\bigr). 
\end{equation} 
Thus, plotting $L^{2\beta/\nu}\langle T \rangle$ against $\tilde{g} \equiv  L^{1/\nu}(g-g_{\text{c}})/\gc$ should give curves that are independent of $L$ and so ``collapse'' onto each other. One can use this curve collapse to fix the values of $\gc$, $\beta$ and $\nu$ directly from data. Following the approach of \cite{Bhattacharjee:2001}, reviewed in Appendix \ref{app:collapse}, one can automate this by defining a measure of how well the curves collapse and then numerically minimizing the measure as a function of $(\gc,\beta,\nu)$.

The data for this procedure is generated by finding L-CNNs which approximate the ground-state wavefunction for $g\in[0.74,0.78]$ in steps of $0.001$ and for lattice sizes $L=4,\dots,12$, and then computing the lattice average of the expectation value of a `t Hooft string of length $L/2$ in the $x$ direction. The raw data is shown in the left plot of Figure \ref{fig:tHooft_collapse}. We note that these curves do not collapse on each other for the range of couplings considered. Following the procedure outlined in Appendix \ref{app:collapse}, we find the values of $\gc$, $\beta$ and $\nu$ that provide the best collapse across the range $\tilde{g}\in[-1,1]$ for $L=8,10,12$,\footnote{Since our finite-size scaling ansatz includes only leading-order corrections, we have used only the three largest lattice sizes to compute the fits -- these should be least affected by subleading corrections in the inverse lattice size.} giving
\begin{equation}\label{eq:our_exponents}
    \gc = 0.7546(8),\qquad\beta = 0.326(4),\qquad\nu = 0.630(3).
\end{equation}

Here, the errors are computed by bootstrapping the fits over the original data. Using these values, in the right plot of Figure \ref{fig:tHooft_collapse} we show that the curves of $L^{2\beta/\nu}\langle T \rangle$ do indeed collapse onto each other as expected. The collapse for $L=6,\dots,12$ is excellent. The collapse holds even for $L=4$, which will have larger corrections in $L^{-1}$ not captured by the leading-order analysis. These results predict that the conformal dimensions of the $\sigma$ and $\epsilon$ primary operators in the critical $2+1$-d Ising model are $\Delta_\sigma = 0.518(4)$ and $\Delta_\epsilon = 1.412(8)$.

We can compare our results with previous calculations of the critical exponents. Since the scaling dimensions of operators are  identical for Ising and Ising$^{\ast}$ CFT, our results should match the ones computed for all systems in the Ising universality class. The three principal theoretical methods of computing critical exponents are Monte Carlo study of lattice systems, perturbative RG flow and the conformal bootstrap. The conformal bootstrap computations are the most precise, but all three techniques are in agreement with each other. From the conformal bootstrap program, state-of-the-art bounds on conformal dimensions yield $\Delta_\sigma = 0.5181489(10)$ and $\Delta_\epsilon = 1.412625(10)$~\cite{Kos:2016ysd,Simmons-Duffin:2016wlq},
which in turn fix
\begin{equation}
    \beta_{\text{CB}} = 0.326419(3), \qquad \nu_\text{CB} = 0.629971(4).
\end{equation}
The most precise Monte Carlo results for exponents of the $2+1$-d Ising model are due to Ferrenberg et al.~\cite{Xu:2018hwn}, extending previous results of Hasenbusch~\cite{10.1103/physrevb.82.174433}, leading to $\nu_\text{MC} = 0.62960(15)$ and $\beta_\text{MC} = 0.32620(31)$.

\begin{figure*}
\hfill{}
\includegraphics[trim={32.10236pt 0 32.10236pt 0}]{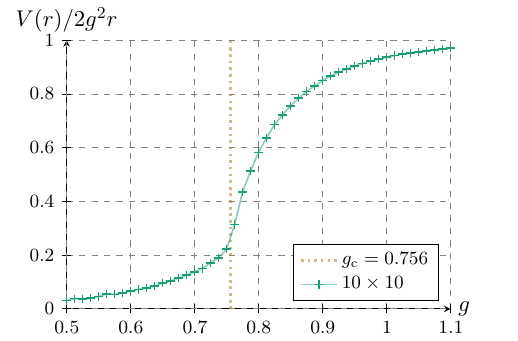}%
\hfill{}
\includegraphics[trim={32.10236pt 0 32.10236pt 0}]{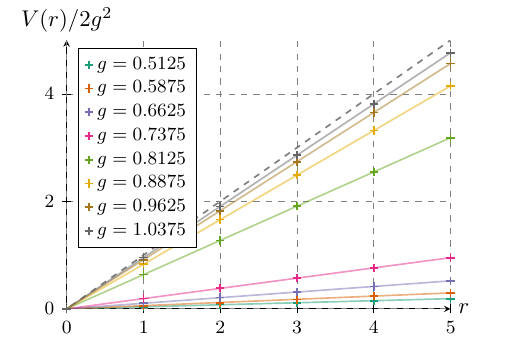}%
\hfill{}
\caption{The left figure shows the linear potential $V(r)$ for $\bZ_2$ gauge theory between two static charges on the lattice divided by its expected strong-coupling limit, $2g^2r$. The charges are placed at $(0,0)$ and $(r,0)$ for $r\in\{1,\dots,5\}$ on a $10\times10$ lattice. The coefficient for each $g$ is computed by fitting a linear function with zero intercept. The right figure shows the potential energy, $V(r) \equiv E_q-E_0$, defined as the difference between the ground-state energy without static charges, $E_0$, and the energy $E_q$ with charges a distance $r$ apart on a $10\times10$ lattice for various values of the coupling. The lines are linear fits of the form $V(r) \propto r$. The $r=0$ data corresponds to the state without charges. From the form of the Hamiltonian \eqref{eq:hamiltonian_Zn}, in the strong-coupling limit, one expects the potential to approach $2g^2 r$, so we have divided by $2g^2$ to see this. We have also plotted this behaviour as the dashed diagonal black line. In all cases, after inserting the static charges, the networks were trained for a further 200 iterations before computing $\langle W_r^\dagger H W_r \rangle_\theta$ (though we observed negligible difference between this and using the untwisted sector ground state).}\label{fig:Z2_charge_potential_slopes}
\end{figure*}

These results are summarised in Table \ref{tab:ising_exponents}. We see that our estimates for $\beta$ and $\nu$ are in excellent agreement with high-precision conformal bootstrap and Monte Carlo values. 
The value of the critical coupling determined by our scaling collapse analysis, $\gc = 0.7546(8)$, is within the $95\%$ confidence interval (two standard deviations) of the value $\gc = 0.757051$ determined by Monte Carlo calculations \cite{PhysRevE.66.066110}. This small deviation can be attributed to the corrections from irrelevant operators in the conformal field theory, which shift the computed value of the critical coupling from the exact value. It is also important to note that our analysis considers only the leading-order effects in the inverse lattice size $L^{-1}$, whereas the Monte Carlo results incorporate subleading corrections, which may further explain the discrepancy in the values of $\gc$.

\begin{table}[h]
	\centering
	\begin{tabular}{c c c} 
		\toprule
		Method & $\nu$ & $\beta$ \tabularnewline
		\midrule
		\midrule
		NNQS & 0.630(3) & 0.326(4) \tabularnewline
 Conformal Bootstrap \cite{Simmons-Duffin:2016wlq} & 0.629971(4) & 0.326419(3) \tabularnewline
 Monte Carlo \cite{10.1103/physreve.97.043301} & 0.62960(15) & 0.32620(31) \tabularnewline
 $\epsilon$ expansion  \cite{Guida1998} & 0.6305(25) & 0.3265(15) \tabularnewline
 Large $N$ \cite{Guida1998} & 0.6304(13) & 0.3258(14) \tabularnewline
		\bottomrule
	\end{tabular}
	\caption{Selected theoretical determinations of $2+1$-d Ising critical exponents.}
	\label{tab:ising_exponents}
\end{table}

\subsubsection{Critical coupling from BST extrapolation}\label{sec:BST_coupling}

So far, we have seen that both the order and disorder parameters display changes in behaviour in the vicinity of critical coupling predicted by Monte Carlo studies of the dual spin system, and that curve collapse can be used to identify the critical coupling and exponents. Here, we provide an alternative estimate for the critical coupling directly from our results. 

Our curve collapse approach relied on the leading-order corrections from finite-size scaling theory. In fact, one can do better than this by taking into account subleading corrections in $L^{-1}$ and extrapolating to the continuum limit. One way to do this is via BST extrapolation~\cite{10.1007/BF02165234, 10.1088/0305-4470/20/8/032, 10.1088/0305-4470/21/11/019}, a technique which uses rational functions to accelerate convergence. In practice, one looks at the pairs $(L,g_\text{cross})$, where the coupling $g_\text{cross}$ is where the curve for lattice size $L$ and the next largest lattice size cross. One then follows an iterative procedure to find the continuum critical coupling, as outlined in Appendix \ref{app:BST}. 

Applying this to the crossing points of $\langle T \rangle$ for a `t Hooft string operator of length $L/2$ gives the estimate\footnote{The error solely from BST extrapolation can be estimated from $\varepsilon_{N-1}^{(0)}$ (see Equation \eqref{eq:BST_error}). The optimal choice of $\omega$ leads to $\varepsilon_{N-1}^{(0)}\sim {10}^{-11}$, implying that the extrapolation has converged well.}
\begin{equation}
    \gc = 0.756431,
\end{equation}
where the error solely due to convergence of the BST procedure is negligible. 
This is slightly larger than the coupling we found via curve collapse, though a small difference is not surprising -- curve collapse considers only leading-order corrections in $L^{-1}$ while BST extrapolation can also account for higher-order corrections, including an $L$-dependent shift of $\gc$.

\subsection{Potential energy between charges}

By using the ground-state wavefunctions obtained after training, we can compute the expectation value of the Wilson string to find the potential between charges on the lattice. Since the gauge group is $\mathbb{Z}_{2}$, the possible charges are $\{0,1\}$. The Wilson string stretching between two non-zero charges lying along the $x$ direction of length $r$ is simply a product of $Z$ (clock) operators 
\begin{equation}
W_{r}=Z_{x,\mu}Z_{x+\hat{x},\mu}\dots Z_{x+r\hat{x},\mu}.
\end{equation}
At small couplings $g\ll1$, the theory is deconfined and the potential per unit length of the Wilson string vanishes. On the other hand, for large values of the coupling, the potential energy can be obtained by setting $N=2$ in \eqref{eq:potential_energy}:
\begin{equation}
    \langle W_{r}^{\dagger}HW_{r}\rangle_0 - \langle H\rangle_0 =2 g^2 r. 
\end{equation}
The expected linear dependence of the potential energy as a function of the distance between charges is confirmed on the right plot in Figure \ref{fig:Z2_charge_potential_slopes}. The potential energy per unit length increases rapidly near the phase transition and saturates to $2g^2$ at strong coupling, as can be seen from the left plot of Figure \ref{fig:Z2_charge_potential_slopes}.

\section{\texorpdfstring{$\mathbb{Z}_3$}{Z3} gauge theory in \texorpdfstring{$2+1$}{2+1} dimensions}\label{sec:Z3}

We now turn to studying $\mathbb{Z}_3$ gauge theory. This theory is expected to have a first-order phase transition separating the deconfined and confined phase~\cite{Horn:1979fy,Kogut:1979wt}. For a first-order transition, the spectral gap does not vanish even in the thermodynamic limit, and therefore the correlation length is always finite. Since the ground state of the system changes abruptly at the point of transition, physical observables such as the energy and its derivatives as a function of coupling display kinks or discontinuities. For finite lattice sizes, the ordered and disordered phases coexist, smoothing out these features. 

By computing the energy and its derivatives, we can determine the critical coupling $\gc$ at which the transition occurs. The computation of other observables such as the Creutz ratio, `t Hooft string operator and the potential energy between charges provides insight into the deconfined and confined phases. The theory is in a deconfined phase below the critical coupling, with perimeter-law decay for Wilson loops and a broken $\mathbb{Z}_3^{(1)}$ one-form symmetry. In the confined phase, the 
Wilson loops obey an area-law decay and the $\mathbb{Z}_3^{(1)}$ one-form symmetry is preserved. As with $\bZ_2$, the expectation value of the `t Hooft string operators become constant due to condensation of magnetic monopoles. 

\subsection{Ground-state energies}

\begin{figure}
\includegraphics{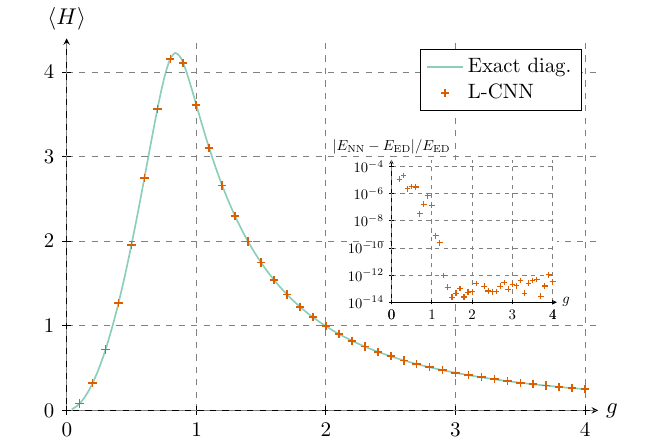}

\caption{Expectation value of $\langle H\rangle$ for $\protect\bZ_{3}$ gauge theory \eqref{eq:hamiltonian_Zn} on a $2\times2$ lattice as a function of coupling $g$. The ground state was computed using an L-CNN with six \textbf{L-CB} layers with $(N_\text{out},K)=(3,1)$. The networks were trained for up to 100 iterations with $4096$ samples. We trained the networks from larger to smaller values of the coupling $g$, using the previously trained network as the starting point. The solid teal line is the exact diagonalization result, computed using NetKet. The inset shows the fractional difference between the ground state energies calculated using the neural network and exact diagonalization.}
\label{fig:energy_Z3_2x2_exact}
\end{figure}

\begin{figure}
\includegraphics{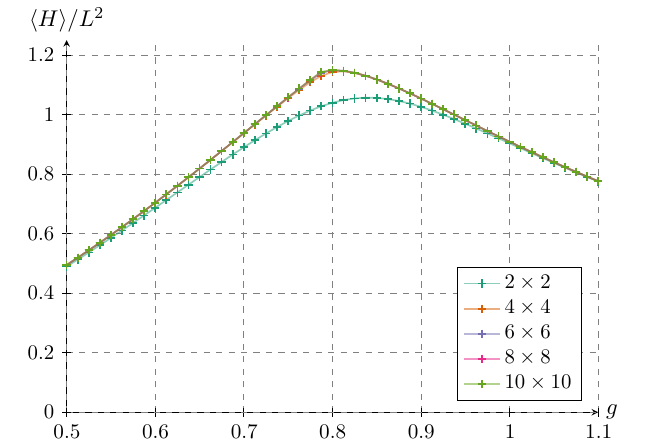}

\caption{Expectation value $\langle H\rangle$ for $\protect\bZ_{3}$ gauge theory for varying lattice sizes as a function of coupling $g$, focused on the region around the critical coupling. The networks were constructed from $\lceil \log_2 N_\text{sites}\rceil$ \textbf{L-CB} layers with $(N_\text{out},K) = (4,2)$, and trained for up to 1000 iterations with 4096 samples. Training was stopped early when the variance of the energy per site stabilised to a value of $1.25\times{10}^{-4}$ or less. The expectation values were then calculated using $32\times4096$ Monte Carlo samples. We trained the networks from larger to smaller values of the coupling $g$, using the previously trained network as the starting point.}
\label{fig:Z3_LxL_energy}
\end{figure}

As a test of our approach, we compare it with previous work using tensor networks to numerically investigate the ground state of a pure $\bZ_{3}$ lattice gauge theory in $2+1$ dimensions~\cite{Emonts:2020drm}. In Figure \ref{fig:energy_Z3_2x2_exact}, we plot the expectation value of the lattice Hamiltonian \eqref{eq:hamiltonian_Zn} in the ground state, i.e.~the ground state energy, as a function of coupling $g$ for a $2\times2$ lattice. We see excellent agreement with the energy calculated by exact diagonalization, and find more accurate results than those recently obtained using tensor networks~\cite{Emonts:2020drm}.

In Figure \ref{fig:Z3_LxL_energy}, we show the ground-state energy on lattices of size $L\in\{2,4,6,8,10\}$ as a function of coupling in the region $g\in[0.5,1.1]$. Compared with the results for $\bZ_2$ in Figure \ref{fig:Z2_LxL_energy}, the curves for $\bZ_3$ appear to be somewhat sharper at the transition around $g\approx0.8$. As we will see in a moment, this can can be seen more clearly by computing the derivative of the energy with respect to the coupling.

\subsection{First-order phase transition}
The zero-temperature $\bZ_{3}$ theory is expected to undergo a first-order phase transition at some critical coupling $\gc$. There are few works in the literature which determine this coupling directly from the gauge theory. From \cite[Figure 7]{Emonts:2020drm}, the behaviour of the string tension on a $6\times6$ lattice suggests $\gc\in[0.75,0.90]$. However, the string tension data is rather noisy, and the authors do not give an explicit estimate of $\gc$. A somewhat cleaner calculation is given in \cite{Robaina:2020aqh} using an infinite projected entangled-pair state (iPEPS) ansatz. The ground-state energy displays a kink at the critical coupling, and a fit to the area-law coefficient agrees with the location of the kink, giving $\gc=0.818$.\footnote{The critical coupling is given in \cite{Robaina:2020aqh} as $\gc^2 = 1.159$. However, their Hamiltonian is not the same as ours. Specifically, working in a basis where the shift operator is diagonal, they have resummed the electric term in the Hamiltonian and absorbed various constants into the coupling. This results in the identifications $g_\text{there}^2 = \sqrt{3} g_\text{here}^2$ and $\sqrt3 H_\text{there}(g_\text{there}) = H_\text{here}(g_\text{here}) - g_\text{here}^{-2} L^2$.}

\begin{figure}
\includegraphics{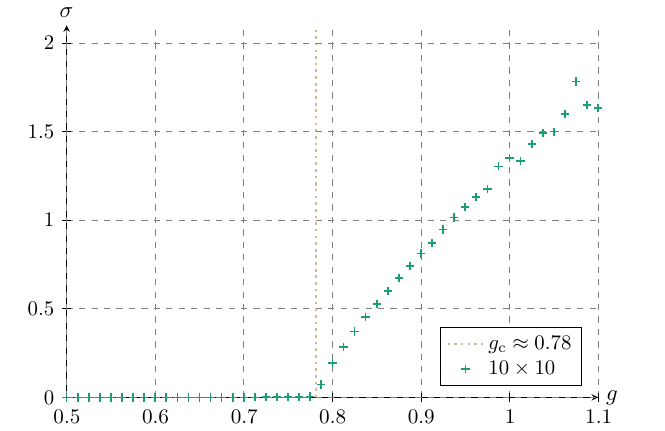}

\caption{String tension estimated from Creutz ratio $\chi_l$ for $l=2$ on a $10\times10$ lattice using $2^{20}\approx 1\text{M}$ Monte Carlo samples. For $g>g_{\text{c}}$, Wilson loops decay rapidly with size. Estimates of $\sigma$ are then challenging due to ratios of very small numbers, as can be seen around $g>0.95$.}
\label{fig:Z3_creutz}
\end{figure}

\begin{figure}
\includegraphics{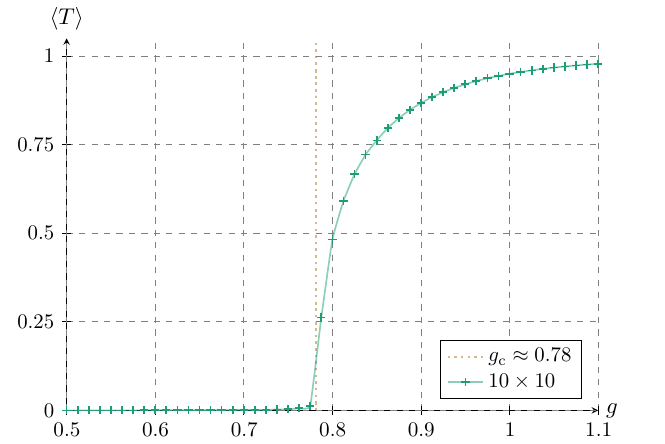}

\caption{The lattice average of the `t~Hooft string operator in $\bZ_3$ gauge theory as a function of coupling $g$ near the critical point for a $10\times10$ lattice. The string operator is of length five in the $x$ direction. The dotted yellow line indicates our identification of the transition at $\gc = 0.781$.}\label{fig:Z3_tHooft_10x10}
\end{figure}

Before locating the critical coupling, we first look for evidence of the confined and deconfined phases. Following a similar discussion in Section \ref{sec:Z2}, in Figure \ref{fig:Z3_creutz}, we plot an estimate of the string tension, computed using the Creutz ratio for $l=2$. At small couplings, the string tension is zero, indicating the presence of an ordered, deconfined phase. At larger couplings, the string tension is non-zero and increasing with coupling, corresponding to a disordered, confined phase. The phase transition appears to occur in the region $g\in[0.7755,0.7875]$. For completeness, we also compute the lattice average of the disorder operator, i.e.~the `t~Hooft string. As for $\bZ_2$, one expects this operator to decay exponentially with the string length in the ordered phase, while it should be independent of length (but coupling dependent) in the disordered phase. Our results in Figure \ref{fig:Z3_tHooft_10x10} confirm these expectations, with the transition region agreeing with that suggested by the string tension.

\begin{figure}
\hfill{}
\includegraphics[trim={21.45244pt 0 32.10236pt 0}]{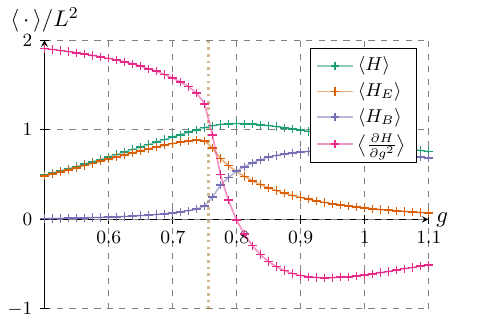}%
\hfill{}
\includegraphics[trim={21.45244pt 0 32.10236pt 0}]{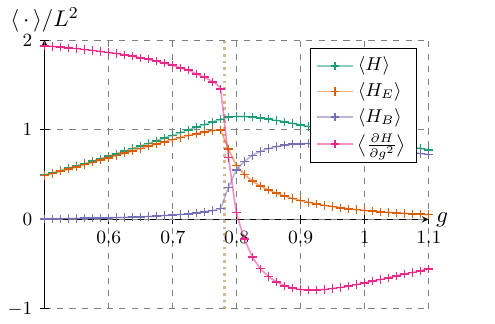}%
\hfill{}
\caption{The left figure shows various expectation values per lattice site for $\bZ_2$ gauge theory computed on a $10\times10$ lattice using an L-CNN as a function of coupling. These include the ground-state energy, the electric and magnetic energies, and the derivative of the energy with respect to $g^2$. We have also labeled the position of the critical coupling $\gc=0.756$ with a dashed gold line as calculated from BST extrapolation in Section \ref{sec:BST_coupling}. The right figure shows the same data for $\bZ_3$ gauge theory, with the transition marked midway between $0.775$ and $0.7875$ at $\gc = 0.781$.}\label{fig:Z2_Z3_E_deriv}
\end{figure}

Note that one would expect a first-order transition to be signalled by discontinuities in order and disorder parameters. For the `t Hooft string, the $\bZ_3$ theory displays a steeper increase than $\bZ_2$ at the transition, however it does not display clear signs of a discontinuity. In fact, this behaviour is not unexpected, and is often referred to as being ``weakly'' first order. First-order transitions are referred to as weakly first order when the (finite) correlation length at the transition is larger than the system size. If the lattice system is smaller than the correlation length $\xi$, it is difficult to differentiate between first-order and continuous transitions, with the expected discontinuities smoothed out, the critical coupling shifted, and finite-size scaling analysis failing to recover trivial critical exponents (i.e.~$\nu=1/d$)~\cite{Iino_2019}.\footnote{For example, the correlation lengths of the two-dimensional $q$-state Potts model were computed by Buffenoir and Wallon~\cite{Buffenoir:1992mz}. The theory has a continuous $\xi\to\infty$) transition for $q=4$, and a first-order transition for $q\geq5$. However, for $q=5,\ldots$ the correlation lengths $\xi\approx 2500,150,50,25,15,10,\ldots$ indicate that one will not see discontinuities in observables for smaller values of $q$ without going to very large lattices.} Indeed, it is known that the spin system dual to $\bZ_3$ gauge theory, the three-state Potts model, has a correlation length $\xi\sim10$~\cite{Fukugita:1989cs}, and so we may need to look at larger lattices to see a clear signal of first-order behaviour.

A first-order quantum phase transition can also be diagnosed by a discontinuity in the first derivative of the energy with respect to the coupling~\cite{Robaina:2020aqh}. Using the Hellmann--Feynman theorem, this derivative can be calculated as
\begin{equation}
\frac{\partial \langle H \rangle_\theta}{\partial g^{2}} \equiv \left< \frac{\partial H}{\partial g^{2}} \right>_\theta = g^{-2}\langle\Psi_\theta|H_{E}-H_{B}|\Psi_\theta\rangle.
\end{equation}
In Figure \ref{fig:Z2_Z3_E_deriv}, we plot the derivative of the energy per lattice site as a function of coupling for both $\bZ_2$ and $\bZ_3$ on a $10\times10$ lattice. In the case of $\bZ_2$, which is known to have a continuous phase transition, the transition region near to the critical coupling ($\gc\approx0.756$) shows a smooth decrease. For $\bZ_3$, the derivative instead shows a steep decrease in the region $[0.775,0.7875]$, indicating a first-order transition. This is also apparent in the curves for $\langle H_B \rangle$, the magnetic term in the Hamiltonian, which display a sharper, more sudden increase for $\bZ_3$.\footnote{Note that $H_B / L^2$ is effectively one minus the single-plaquette Wilson loop $W_\Box$, so we have not plotted this separately.} From this, and the order and disorder parameters, we identify the critical coupling to be 
\begin{equation}
    \gc = 0.781(6).
\end{equation}
where the error corresponds to the half-width of the interval $[0.775,0.7875]$. If one plots the same expectations for varying lattice size, one sees that the curve for $\bZ_2$ does not change much between $L=6$ and $L=10$, suggesting that it is not becoming discontinuous in the continuum limit. The curves for $\bZ_3$ instead show the pattern of getting steeper and steeper with increasing $L$, exactly as one would expect for a first-order phase transition.

We can also  check whether our variational state has correctly learned the symmetries of the ground state. For example, we know that the true ground state should be invariant under lattice translations, rotations and reflections, and charge conjugation. L-CNNs are manifestly translationally invariant, so this symmetry is exact. The remaining symmetries will be learned as part of the training process, and so they will approximate. One can check this by comparing the amplitude $\Psi(\mathcal{U})$ with $\Psi(\mathcal{U'})$, where $\mathcal{U}'$ is related to $\mathcal{U}$ by one of the symmetry transformations. For example, charge conjugation acts by sending all link variables to their conjugates, $\mathcal{U}'=\mathcal{U}^\dagger$. Charge conjugation invariance then implies that expectation values of Wilson loops for the ground state should be \emph{real}.\footnote{In fact, since the Hamiltonian commutes with time reversal, if the ground-state is unique, one can always pick the ground-state to be real (see, for instance, \cite[Equation 6.4]{Cotler:2023lem}). Then, if it is also charge-conjugation invariant, one also has $\Psi(\mathcal{U}) = \Psi(\mathcal{U}^\dagger)$. This symmetry could be imposed in the network architecture, though we have not done so in this work.} Denoting the charge conjugation operator by $C$, this follows from
\begin{equation}
    \langle W \rangle_0 = \langle\Psi_0|C^\dagger W C|\Psi_0\rangle = \langle W \rangle_0^\dagger,
\end{equation}
where we have used $C|\Psi_0\rangle = |\Psi_0\rangle$ for the ground state. Again, we can check to what accuracy this holds for our trained networks by comparing the size of the imaginary part of the expectation value with its magnitude. On a $10\times10$ lattice, evaluating the lattice average of the single-plaquette Wilson loop operator for our NNQSs with $2^{18}\approx 250\text{k}$ samples gives values of $\im \langle W_\Box \rangle_\theta / |\langle W_\Box \rangle_\theta|$ in the range ${10}^{-6}$ to ${10}^{-4}$ for $g\in[0.5,1.1]$. From this, we conclude that our networks have learned the charge conjugation symmetry to high accuracy.

\subsection{Static charges}
Finally, we turn to the potential between static charges in the $\mathbb{Z}_{3}$ theory, computed using the expectation value of the Hamiltonian conjugated by a Wilson string. In this case, the possible charges are $\{0,1,2\}$ and thus there are two distinct Wilson strings. The following Wilson string operator places the charges $1$ at $x$ and $2$ at $x+r\hat{x}$, a distance $r$ apart along the $x$-direction:
\begin{equation}
W_{r}=Q_{x,\mu}Q_{x+\hat{x},\mu}\dots Q_{x+r\hat{x},\mu}.
\end{equation}
The square of this operator $W_{r}^{2} = W_{r}^{\dagger}$ places charges $2$ at $x$ and $1$ at $x+r\hat{x}$, so the potential energy associated to it is identical to that of $W_{r}$.

\begin{figure}
\hfill{}
\includegraphics[trim={32.10236pt 0 32.10236pt 0}]{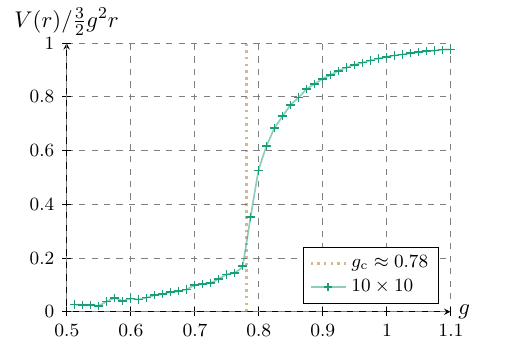}%
\hfill{}
\includegraphics[trim={32.10236pt 0 32.10236pt 0}]{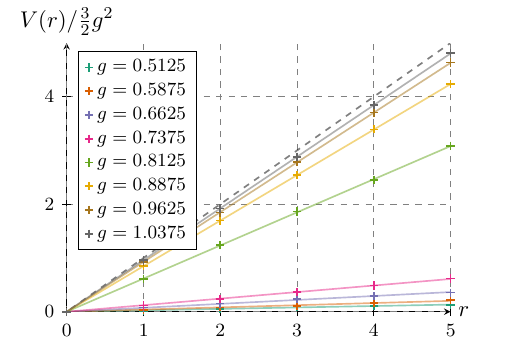}%
\hfill{}
\caption{The left figure shows the linear potential $V(r)$ for $\bZ_3$ gauge theory between two static charges on the lattice divided by its expected strong-coupling limit, $\frac32 g^2$. The charges are placed at $(0,0)$ and $(r,0)$ for $r\in\{1,\dots,5\}$ on a $10\times10$ lattice. The coefficient for each $g$ is computed by fitting a linear function with zero intercept. The right figure shows the potential energy, $V(r) \equiv E_q-E_0$, defined as the difference between the ground-state energy without static charges, $E_0$, and the energy $E_q$ with charges a distance $r$ apart on a $10\times10$ lattice for various values of the coupling. The lines are linear fits of the form $V(r) \propto r$. The $r=0$ data corresponds to the state without charges. From the form of the Hamiltonian \eqref{eq:hamiltonian_Zn}, in the strong-coupling limit, one expects the potential to approach $\tfrac{3}{2}g^2 r$, so we have divided by $\tfrac{3}{2}g^2$ to see this. We have also plotted this behaviour as the dashed diagonal black line. In all cases, after inserting the static charges, the networks were trained for a further 200 iterations before computing $\langle W_r^\dagger H W_r \rangle$.}\label{fig:Z3_charge_potential_slopes}
\end{figure}

The theory is deconfined at very small values of the coupling, and so the potential per unit length should vanish. Since there is a first-order phase transition for $\mathbb{Z}_{3}$, the potential energy per unit distance should increase drastically near the phase transition, with this behaviour becoming singular in the thermodynamic limit. For very large values of $g$, the potential energy can be estimated by setting $N=3$ in \eqref{eq:potential_energy} 
\begin{equation}
   V(r) \approx \langle W_{r}^{\dagger}HW_{r}\rangle_0 - \langle H\rangle_0 = \frac{3}{2} g^2 r. 
\end{equation}
The dramatic increase in the potential at the phase transition and the predicted weak-/strong-coupling behaviour is demonstrated in the left plot of Figure \ref{fig:Z3_charge_potential_slopes}, while the linear dependence of the potential energy on the distance between charges is illustrated in the right plot. Comparing these with the analogous results for $\bZ_2$ in Figure \ref{fig:Z2_charge_potential_slopes}, we see that both display potentials that increase linearly with distance, with the increase at the phase transition being smooth for $\bZ_2$ and much sharper for $\bZ_3$, in agreement with their continuous and first-order nature respectively.

\subsection*{Acknowledgements}

We thank Miranda Cheng, Edward Mazenc, Matija Medvidović and Semon Rezchikov for stimulating discussions and useful insights, and Filippo Vicentini for patiently answering the authors' questions about NetKet. The work of A.~Apte is supported by the Data Science Institute at the University of Chicago. A.~Ashmore is supported by the European Union’s Horizon 2020 research and innovation program under the Marie Sk\l{}odowska-Curie grant agreement No.~838776, and acknowledges previous support from NSF Grant PHY-2014195 and the Kadanoff Center for Theoretical Physics. Part of this work was carried out at the Aspen Center for Physics, which is supported by NSF grant PHY-2210452. C.~C\'ordova is partially supported by the US Department of Energy Grant 5-29073 and the Sloan Foundation. C.~C\'ordova and T.-C.~Huang also acknowledge support from the Simons Collaboration on Global Categorical Symmetries. We are grateful to the University of Chicago’s Research Computing Center and AI+Science Research Initiative for assistance with the calculations carried out in this work.

\appendix

\section{Implementation of Wilson loop operators}\label{app:WL}

In NetKet, Wilson loop operators can be constructed either by chaining together products of \texttt{LocalOperator}'s, or by constructing a custom operator. The first approach often runs into memory problems, as large Wilson loops need products of many local operators (e.g.~an $8\times8$ Wilson loop will need a product of 32 local operators). Instead, we constructed a custom operator. The implementation is simplified by the fact that we are interested only in the expectation value of the Wilson loop operators, and not in their gradients with respect to the network parameters (which would require a more complicated implementation).
Given a state $|\Psi\rangle$, one can compute the expectation value of an operator $\mathcal{O}$ via
\begin{equation}
\langle\mathcal{O}\rangle=\frac{\langle\Psi|\mathcal{O}|\Psi\rangle}{\langle\Psi|\Psi\rangle}=\sum_{\mathcal{U}}\frac{|\Psi(\mathcal{U})|^{2}}{\langle\Psi|\Psi\rangle}\left(\sum_{\mathcal{U}'}\frac{\langle\mathcal{U}|\mathcal{O}|\mathcal{U}'\rangle\langle\mathcal{U}'|\Psi\rangle}{\langle\mathcal{U}|\Psi\rangle}\right)=\mathbb{E}_{\mathcal{U}\sim|\Psi(\mathcal{U})|^{2}}\bigl[\mathcal{O}_{\text{loc}}(\mathcal{U})\bigr],
\end{equation}
where $\mathcal{U}$ and $\mathcal{U}'$ are gauge field configurations, $\mathbb{E}_{\mathcal{U}\sim|\Psi(\mathcal{U})|^{2}}$ means take the expectation value over $\mathcal{U}$, with $\mathcal{U}$ distributed according to the PDF $|\Psi(\mathcal{U})|^{2}$, and we have defined the ``local estimator''
\begin{equation}
\mathcal{O}_{\text{loc}}(\mathcal{U})
=\frac{\langle\mathcal{U}|\mathcal{O}|\Psi\rangle}{\langle\mathcal{U}|\Psi\rangle}
=\sum_{\mathcal{U}'}\frac{\langle\mathcal{U}|\mathcal{O}|\mathcal{U}'\rangle\langle\mathcal{U}'|\Psi\rangle}{\langle\mathcal{U}|\Psi\rangle}
=\sum_{\mathcal{U}'}\langle\mathcal{U}|\mathcal{O}|\mathcal{U}'\rangle \frac{\Psi(\mathcal{U}')}{\Psi(\mathcal{U})} .
\end{equation}
In order to evaluate this expectation value via variational Monte Carlo, one needs the following:
\begin{enumerate}

\item Some way of restricting the sum over $\mathcal{U}$ to finitely many representative values by sampling the PDF, $|\Psi(\mathcal{U})|^{2}$. This is provided by the Monte Carlo variational state interface of NetKet.

\item A method to take the samples $\{\mathcal{U}\}$ and compute the ``connected elements'', i.e.~those for which the matrix element $\langle\mathcal{U}|\mathcal{O}|\mathcal{U}'\rangle\neq0$ is non-zero. This means that for each $\mathcal{U}$, one does not have to sum over all of the sampled configurations again, but can instead restrict to a (usually much) smaller set of configurations, labeled by $\{\mathcal{U}'\}$. One then also needs to compute the matrix elements themselves.

\item A calculation of the local estimators $\mathcal{O}_{\text{loc}}(\mathcal{U})$ given the matrix elements, the sets $\{\mathcal{U}\}$ and $\{\mathcal{U}'\}$, and the state $|\Psi\rangle$. Again, this implementation is provided by NetKet.

\item The statistical average of the local energies weighted by $|\Psi(\mathcal{U})|^{2}$.

\end{enumerate}
For a single clock operator $Q_{l}$, it is simple to find the connected elements $\{\mathcal{U}'\}$ and the matrix elements $\langle\mathcal{U}|\mathcal{O}|\mathcal{U}'\rangle$. Given a configuration of link variables $|\mathcal{U}\rangle=|q\rangle_{1}\otimes\dots\otimes|q\rangle_{l}\otimes\dots$ on the lattice, $Q_{l}$ acts as
\begin{equation}
Q_{l}|\mathcal{U}\rangle=\ee^{2\pi\ii q_{l}/N}|\mathcal{U}\rangle.
\end{equation}
Since we are working in a basis which diagonalises the action of $Q_{l}$, this preserves the configuration up to a phase, so that the connected set $\{\mathcal{U}'\}$ has only one element, namely $\mathcal{U}$ itself. That is
\begin{equation}
\langle\mathcal{U}'|Q_{l}|\mathcal{U}\rangle=\ee^{2\pi\ii q_{l}/N}\delta_{\mathcal{U},\mathcal{U}'}.
\end{equation}
The same holds for products of clock operators, such as those used for constructing Wilson loop operators. Using this, it is relatively straightforward to implement a custom Wilson loop operator in NetKet.

\section{Exponents from data collapse}\label{app:collapse}

Finite-size scaling is one of the key pillars in the theory of critical phenomena~\cite{Stanley:1999}. Consider a system of characteristic length scale $L$ close to the critical point $\gc$. The behaviour of a physical quantity $m$ in the vicinity of the transition can be captured by a scaling relation~\cite{Widom:1974}
\begin{equation}\label{eq:scale}
    m (t,L) = L^{d}f(t/L^c),
\end{equation}
where $t=(g-\gc)/\gc$ is the reduced coupling, and depending on the physical system $m$ may refer to quantities such as specific heat, magnetization, or in our case the expectation value of the `t Hooft string operator. The remarkable aspect of this relation is that it holds for different physical systems in the same universality class. The scaling relation predicts that when the data for $m L^{-d}$ is plotted against $t L^{-c}$, the curves for different system sizes (and even for different materials belonging to the same universality class) should collapse onto a single curve. See, for example, the first figure in \cite{Stanley:1999} for a striking illustration of curve collapse in magnetization data for five different materials belonging to the $2+1$-d Heisenberg universality class. 

The scaling relation can be used to extract the values of the exponents $c$ and $d$, and the critical coupling $\gc$ based on numerical data for $m(t,L)$. Given a measure for the data collapse, numerical minimization techniques can be used to automatically search for the exponents. Consider a collection of data for a set of lengths $\{L_j\}$. The tabulated values of $m$ and $t$ can be represented as $\{m_{ij}\}$ and $\{t_{ij}\}$ respectively, where $ij$ labels the $i$'th value of $t$ for length $L_j$. The central obstacle to overcome in defining a measure of collapse is the absence of the knowledge of the scaling function $f$. Closely following \cite{Bhattacharjee:2001}, we can use a polynomial interpolation based on any length $L_{p} \in \{L_j\}$ to determine $f$ and compute the deviation from this curve. Since any length can be used for this purpose, we repeat the procedure for all lengths and average to obtain a measure of collapse:
\begin{equation}\label{eq:collapse_errors}
    P(c,d,g_\text{c}) = \frac{1}{N_{\textrm{over}}} \sum_{p} \sum_{i,j} \Bigl| m_{ij} L_{j}^{-d}- \mathcal{E}_{p}(t_{ij} L_{j}^{-c})\Bigr|,
\end{equation}
where $\mathcal{E}_{p}(x)$ is the interpolation function based on $L_p$. The sum over $i$ is performed only in the overlapping region to avoid extrapolation, and $N_{\textrm{over}}$ is the total number of points used to compute $P$. We use a cubic spline interpolation scheme to find $\mathcal{E}_{p}(x)$. Since $P \geq 0$ and vanishes only for perfect data collapse, minimization of $P$ can be used to extract the optimal values of the parameters. Since each of the terms that appears in the sum is non-negative, we can use the method of least-squares \cite{Laplace1812} to compute the optimal values of $(c,d,\gc)$ that lead to the best-fit. The uncertainties in the computed values are estimated by using statistical bootstrapping \cite{Efron1979}, wherein the computation of optimal parameters is carried out multiple times using data that is re-sampled from the original dataset. This method of computing uncertainties in critical exponents is routinely employed in the Monte Carlo simulations of statistical systems \cite{Beach2005, Suwa2016}. 

\section{BST extrapolation}\label{app:BST}

Finite-size scaling analysis has long been used to extrapolate finite-size lattice data to the thermodynamic limit, $L\to\infty$. In this appendix, we give a brief review of one such technique for extrapolating a sequence of finite-size data which empirically has been found to converge rapidly. This is known as BST extrapolation~\cite{10.1007/BF02165234, 10.1088/0305-4470/20/8/032, 10.1088/0305-4470/21/11/019}.

The general set-up is as follows. Certain continuum observables, such as the critical coupling $g_{\text{c}}$, admit an asymptotic form
\begin{equation}
	g_{\text{c}}(L)=g_{\text{c}}+a_{1}L^{-\omega_{1}}+a_{2}L^{-\omega_{2}}+\dots,
\end{equation}
where $g_{\text{c}}(L)$ is computed via a finite lattice simulation and $g_{\text{c}}$ is the continuum critical coupling that one is interested in predicting to high accuracy.

Let $h_{i}$, $i=0,\dots,N$, denote a sequence of positive numbers which converges to zero for $i\to\infty$. In practice, one often takes the $h_{i}$ to be given by the inverse lattice size. As an example, consider even lattices with $h_{i}=(1/2,1/4,1/6,1/8)$ with $N=3$ and $i=0,\dots,3$. Assume that the continuum observable of interest is $T$, which has a finite-size expansion
\begin{equation}
	T(h)=T+a_{1}h^{\omega_{1}}+a_{2}h^{\omega_{2}}+\dots,\label{eq:T(h)}
\end{equation}
where $\omega_{2}>\omega_{1}$, and so on. One then computes the following quantities
\[
\begin{array}{ccccccccc}
	m=0 & T_{0}^{(0)} &  & T_{0}^{(1)} &  & T_{0}^{(2)} &  & T_{0}^{(3)} & \phantom{i=0}\\
	m=1 &  & T_{1}^{(0)} &  & T_{1}^{(1)} &  & T_{1}^{(2)} &  & \phantom{i=1}\\
	m=2 &  &  & T_{2}^{(0)} &  & T_{2}^{(1)} &  &  & \phantom{i=2}\\
	m=3 &  &  &  & T_{3}^{(0)} &  &  &  & \phantom{i=3}
\end{array}
\]
using the iterative rules
\begin{equation}
	\begin{aligned}T_{-1}^{(i)} & =0,\\
		T_{0}^{(i)} & =T(h_{i}),\\
		T_{m}^{(i)} & =T_{m-1}^{(i+1)}+(T_{m-1}^{(i+1)}-T_{m-1}^{(i)})\left[\left(\frac{h_{i}}{h_{i+m}}\right)^{\omega}\left(1-\frac{T_{m-1}^{(i+1)}-T_{m-1}^{(i)}}{T_{m-1}^{(i+1)}-T_{m-2}^{(i+1)}}\right)-1\right]^{-1},
	\end{aligned}
	\label{eq:BST}
\end{equation}
where the final line is evaluated for $m\geq1$. These rules come from approximating $T(h)$ via a sequence of rational functions.

The point of this is that the original sequence of finite-size data $T_{0}^{(i)}$ is expected to converge slowly to the continuum limit, with $T_{0}^{(3)}$ in our example being the best estimate that one can obtain from the four lattice sizes considered. Comparing with the expansion of $T(h)$ in \eqref{eq:T(h)}, this estimate obeys
\begin{equation}
	T_{0}^{(3)}=T+\mathcal{O}(h_{3}^{\omega_{1}}).\label{eq:first_correction}
\end{equation}
Using the rules above to calculate the $m=1$ row, our new best estimate will be $T_{1}^{(2)}$. It then follows straightforwardly from the form of \eqref{eq:BST} that, upon picking $\omega=\omega_{1}$, this estimate obeys
\begin{equation}
	T_{1}^{(2)}=T+\mathcal{O}(h_{2}^{\eta}),
\end{equation}
where $\eta=\min(\omega_{2},2\omega_{1})$. Provided that the $h_{i}$ are chosen appropriately (i.e.~close enough), this correction term will be smaller than the naive one in \eqref{eq:first_correction}. In other words, in one step of the BST algorithm, one can remove the leading-order correction in $T(h)$. One then iterates to eventually find $T_{3}^{(0)}$, which gives the ``best'' estimate of $T$.

It then remains to choose an appropriate value of $\omega$ given the data $T(h_{i})$. In practice, this is done by calculating
\begin{equation}\label{eq:BST_error}
	\varepsilon_{m}^{(i)}=2|T_{m}^{(i+1)}-T_{m}^{(i)}|.
\end{equation}
Taking $i=0$ and $m=N-1$, $\varepsilon_{N-1}^{(0)}$ measures the difference between the extrapolants in the penultimate row. One should then choose $\omega$ to minimise this quantity. The idea behind this is that, in the $N\to\infty$ limit of many data points, one expects $|T_{m}^{(i)}-T|\leq\varepsilon_{m}^{(i)}$, and so minimizing the difference between the penultimate estimates will lead to a prediction $T_{N}^{(0)}$ for $T$ with the smallest extrapolation error.

\bibliographystyle{utphys}
\bibliography{references}

\end{document}